\documentclass[useAMS,usenatbib]{mn2e}
\usepackage{psfrag,graphicx}
\usepackage{color}
\usepackage{txfonts}
\usepackage{breqn}
\usepackage[T1]{fontenc}
\usepackage{ae,aecompl}
\title[Black hole formation in the early universe]
{How realistic UV spectra and X-rays suppress the abundance of direct collapse black holes}
\author[Latif et al.]
  {M. A. Latif, \thanks{Corresponding author: mlatif@astro.physik.uni-goettingen.de}$^1$
  S. Bovino,$^1$
  T. Grassi, $^{2,3}$
  D. R. G. Schleicher, $^1$
  M. Spaans$^4$\\
   $^1$ Institut f\"ur Astrophysik, Georg-August-Universit\"at, Friedrich-Hund-Platz 1, D-37077 G\"ottingen, Germany \\
   $^2$ Centre for Star and Planet Formation, Natural History Museum of Denmark, \O ster Voldgade 5-7, DK-1350 Copenhagen, Denmark\\
   $^3$ Niels Bohr Institute, University of Copenhagen, Juliane Maries Vej 30, DK-2100 Copenhagen, Denmark \\
   $^4$ Kapteyn Astronomical Institute, University of Groningen, P.O. Box 800, 9700 AV Groningen, The Netherlands 
  }
\date{}

\pagerange{\pageref{firstpage}--\pageref{lastpage}} \pubyear{2009}

\def\LaTeX{L\kern-.36em\raise.3ex\hbox{a}\kern-.15em
      T\kern-.1667em\lower.7ex\hbox{E}\kern-.125emX}

\begin{document}

\bibliographystyle{mn2e}

\label{firstpage}

\maketitle
\def\na{NewA}%
          % New~Astronomy
\def\aj{AJ}%
          % Astronomical Journal
\def\araa{ARA\&A}%
          % Annual Review of Astron and Astrophys
\def\apj{ApJ}%
          % Astrophysical Journal
\def\apjl{ApJ}%
          % Astrophysical Journal, Letters
\def\jcap{JCAP}

\def\apjs{ApJS}%
          % Astrophysical Journal, Supplement
\def\ao{Appl.~Opt.}%
          % Applied Optics
\def\apss{Ap\&SS}%
          % Astrophysics and Space Science
\def\aap{A\&A}%
          % Astronomy and Astrophysics
\def\aapr{A\&A~Rev.}%
          % Astronomy and Astrophysics Reviews
\def\aaps{A\&AS}%
          % Astronomy and Astrophysics, Supplement
\def\azh{AZh}%
          % Astronomicheskii Zhurnal
\def\baas{BAAS}%
          % Bulletin of the AAS
\def\jrasc{JRASC}%
          % Journal of the RAS of Canada
\def\memras{MmRAS}%
          % Memoirs of the RAS
\def\mnras{MNRAS}%
          % Monthly Notices of the RAS
\def\pra{Phys.~Rev.~A}%
          % Physical Review A: General Physics
\def\prb{Phys.~Rev.~B}%
          % Physical Review B: Solid State
\def\prc{Phys.~Rev.~C}%
          % Physical Review C
\def\prd{Phys.~Rev.~D}%
          % Physical Review D
\def\pre{Phys.~Rev.~E}%
          % Physical Review E
\def\prl{Phys.~Rev.~Lett.}%
          % Physical Review Letters
\def\pasp{PASP}%
          % Publications of the ASP
\def\pasj{PASJ}%
          % Publications of the ASJ
\def\qjras{QJRAS}%
          % Quarterly Journal of the RAS
\def\skytel{S\&T}%
          % Sky and Telescope
\def\solphys{Sol.~Phys.}%
          % Solar Physics

          % Solar Physics
\def\sovast{Soviet~Ast.}%
          % Soviet Astronomy
\def\ssr{Space~Sci.~Rev.}%
          % Space Science Reviews
\def\zap{ZAp}%
          % Zeitschrift fuer Astrophysik
\def\nat{Nature}%
          % Nature
\def\iaucirc{IAU~Circ.}%
          % IAU Cirulars
\def\aplett{Astrophys.~Lett.}%
          % Astrophysics Letters
\def\apspr{Astrophys.~Space~Phys.~Res.}%
          % Astrophysics Space Physics Research
\def\bain{Bull.~Astron.~Inst.~Netherlands}%
          % Bulletin Astronomical Institute of the Netherlands
\def\fcp{Fund.~Cosmic~Phys.}%
          % Fundamental Cosmic Physics
\def\gca{Geochim.~Cosmochim.~Acta}%
          % Geochimica Cosmochimica Acta
\def\grl{Geophys.~Res.~Lett.}%
          % Geophysics Research Letters
\def\jcp{J.~Chem.~Phys.}%
          % Journal of Chemical Physics
\def\jgr{J.~Geophys.~Res.}%
          % Journal of Geophysics Research
\def\jqsrt{J.~Quant.~Spec.~Radiat.~Transf.}%
          % Journal of Quantitiative Spectroscopy and Radiative Trasfer
\def\memsai{Mem.~Soc.~Astron.~Italiana}%
          % Mem. Societa Astronomica Italiana
\def\nphysa{Nucl.~Phys.~A}%
          % Nuclear Physics A
\def\physrep{Phys.~Rep.}%
          % Physics Reports
\def\physscr{Phys.~Scr}%
          % Physica Scripta
\def\planss{Planet.~Space~Sci.}%
          % Planetary Space Science
\def\procspie{Proc.~SPIE}%
          % Proceedings of the SPIE

% 

% \newcommand{\ch}[1]{\textcolor{red}{\textbf{#1}}}
% \newcommand{\newln}{\\&\quad\quad{}}
% \date{today}

\begin{abstract}
{
Observations of high redshift quasars at $z>6$ indicate that they harbor supermassive black holes (SMBHs) of a billion solar masses. The direct collapse scenario has emerged as the most plausible way to assemble SMBHs. The nurseries for the direct collapse black holes are massive primordial halos illuminated with an intense UV flux emitted by population II (Pop II) stars. In this study, we compute the critical value of such a flux ($J_{21}^{\rm crit}$) for realistic spectra of Pop II stars through three-dimensional cosmological simulations. We derive the dependence of $J_{21}^{\rm crit}$ on the radiation spectra, on variations from halo to halo, and on the impact of X-ray ionization. Our findings show that the value of $J_{21}^{\rm crit}$ is a few times $\rm 10^4$ and only weakly depends on the adopted radiation spectra in the range between $T_{\rm rad}=2 \times 10^4-10^5$ K. For three simulated halos of a few times $\rm 10^{7}$~M$_{\odot}$, $J_{21}^{\rm crit}$ varies from $\rm 2 \times 10^4 - 5 \times 10^4$. The impact of X-ray ionization is almost negligible and within the expected scatter of $J_{21}^{\rm crit}$ for background fluxes of $J_{\rm X,21} \leq 0.1$.
The computed estimates of $J_{21}^{\rm crit}$ have profound implications for the quasar abundance at $z=10$ as it lowers the number density of black holes forming through an isothermal direct collapse by a few orders of magnitude below the observed black holes density. However, the sites with moderate amounts of $\rm H_2$ cooling may still form massive objects sufficient to be compatible with observations. 
% as it reduces the number density of direct collapse black holes by a few orders of magnitude below the observed value, and puts the direct collapse scenario in hot water.
}
\end{abstract}
% context
% aims as it only elevates $J_{21}^{\rm crit}$ for one halo by a factor of two.

%  results

% conclusion 
\begin{keywords}
methods: numerical -- cosmology: theory -- early Universe -- galaxies: formation
\end{keywords}

\section{Introduction}

Observations of quasars at $z > 6$ unfold the presence of supermassive black holes (SMBHs) of $10^9$~M$_{\odot}$ \citep{2003AJ....125.1649F,2006AJ....131.1203F,2010AJ....139..906W,2011Natur.474..616M,2013ApJ...779...24V}. The formation of such massive objects within the first billion years after the Big Bang is an open question. Various scenarios for the assembling of SMBHs have been proposed which include the remnants of the first stars, stellar dynamical process in a dense nuclear cluster and the collapse of a protogalactic gas cloud into a so-called direct collapse black hole (DCBH). The most natural candidates to assemble SMBHs are the stellar mass black holes formed through the collapse of the first generation of stars.

According to the hierarchical scenario of structure formation  population III (Pop III) stars are formed in minihalos of $10^5-10^6$~M$_{\odot}$ at $z\sim 20-30$. The three dimensional cosmological simulations exploring the formation of Pop III stars including their radiation feedback show that they can reach maximum masses of a few hundred solar \citep{Hirano14,Susa14}. The feedback from stellar mass black holes limits the accretion and makes them difficult to grow in the available time \citep{Johnson07,Alvarez09}. However, cyclic episodes of super Eddington accretion \citep{Madau14} or merging of many stellar mass black holes \citep{Haiman04,Tanaka09} may help them to grow faster. Another way could be the formation of a dense nuclear cluster in metal enriched halo which collapses via the stellar dynamical processes \citep{Devecchi09,Lupi14} and provides a black hole seed of a few thousand solar masses. An alternative route could be the formation of heavy seeds of $10^4-10^6$~M$_{\odot}$ through rapid inflow of gas provided that fragmentation remains inhibited \citep{Bromm03,Spaans06,Begelman06,Lodato06,Latifstream14}. Further details about these mechanisms can be found in reviews \citep{2010A&ARv..18..279V,2012arXiv1203.6075H}.

% Yue13,Yue14,VisbalSync14,,,Ferrara014

The direct collapse scenario has gained a lot of interest during the past decade as it provides massive seeds which can later grow at relatively moderate accretion rates to form SMBHs. Numerical simulations show that massive primordial halos of $\rm 10^7-10^8$~M$_{\odot}$ forming at $z\sim 15-20$ collapse monolithically in the absence of $\rm H_{2}$ cooling \citep{Wise08,Regan09,Latif2011b} and are the potential birthplaces of DCBHs. These studies employed a Jeans resolution of 16 cells insufficient to resolve turbulent eddies. Our recent studies show that resolving turbulence mandates a Jeans resolution of $\geq 32$ cells \citep{Latif13a,LatifpopIII13} which also helps in regulating the angular momentum. \cite{Latif13c} show that fragmentation occurs occasionally but does not prevent the formation of massive objects and large accretion rates of 1~M$_{\odot}/yr$ are observed. The presence of strong magnetic fields amplified via the small scale dynamo further aids in suppressing fragmentation \citep{Latifdynamo13,Latif14Mag}.

The most recent high resolution studies following the evolution after the initial collapse demonstrate that massive seeds of $\rm 10^{5}$~M$_{\odot}$ are formed \citep{LatifCharacteristicBH,Regan14}. Theoretical studies suggest that such supermassive stars are the potential embryos of DCBHs \citep{Begelman10,Ball012,Hosokawa13}. The work by \cite{Schleicher13} suggests that for accretion rates $< 0.1$~M$_{\odot}/yr$ supermassive stars (stars of $\rm 10^{3}-10^6$~M$_{\odot}$) are expected to form while for $> 0.1$~M$_{\odot}/yr$ quasi-stars (stars with BHs in their interior) are the expected outcome. \cite{Ferrara014} have computed the initial mass function of DCBHs from merger tree simulations and found that their mass distribution depends whether their progenitors were polluted or remained pristine. 

It is imperative for the feasibility of this scenario that gas in the halo must be of primordial composition and the formation of molecular hydrogen remains inhibited. The latter mandates the presence of an intense UV flux below the Lyman limit (13.6 eV) to suppress the $\rm H_{2}$ formation \citep{Schleicher10,Latif11a}. Th strength of such a UV flux is much higher than expected background UV field \citep{Dijkstra14} and can be achieved in the vicinity of a star forming galaxy \citep{Dijkstra08,Agarwal12} or even in a synchronized pair of halos \citep{VisbalSync14}.

The main pathway for the formation of $\rm H_{2}$ in primordial gas is:
\begin{equation}
%\mathrm{ H + e^{-} \rightarrow H^{-} + \gamma} \\
\mathrm{H + e^{-} \rightarrow H^{-} +}  \gamma
\end{equation}
\begin{equation}
\mathrm{ H + H^{-} \rightarrow  H_{2} + e^-.} 
\label{h21}
\end{equation}
The formation of $\rm H_{2}$ can be quenched either by destroying  $\rm H_{2}$ or $\rm H^{-}$. Photons with energy between 11.2-13.6 eV are absorbed in the Lyman-Werner bands of $\rm H_{2}$ and dissociate it shortly after putting it into an excited state, this is known as the Solomon process. On the other hand,  $\rm H^{-}$ is photo-detached by photons above $0.76$ eV. The reactions for these processes are the following:
\begin{equation}
%\mathrm{ H + e^{-} \rightarrow H^{-} + \gamma} \\
\mathrm{H_{2}} + \gamma_{LW} \mathrm{\rightarrow H + H}
\label{h20}
\end{equation}
\begin{equation}
\mathrm{H^{-}} + \gamma_{0.76} \mathrm{\rightarrow H + e^{-}}\\ 
\label{h2}
\end{equation}
The former process (i.e. equation \ref{h20}) is efficient for stars with $T_{\rm rad}=10^5$ K while the stars with $T_{\rm rad}=10^4$ K are efficient in photo-detachment of $\rm H^-$.

The critical strength of the radiation flux (hereafter, $J_{21}^{\rm crit}$) above which the formation of $\rm H_2$ remains suppressed has been computed both from one-zone and three dimensional simulations \citep{2001ApJ...546..635O,Shang10,2014MNRAS.443.1979L}. The value of $J_{21}^{\rm crit}$ depends on the spectral shape of radiation field. It is commonly presumed that such UV flux is emitted by the first or second generation of stars using idealized spectra with black body radiation temperatures of $\rm 10^{5}$ or $\rm 10^{4}$ K. The value of $J_{21}^{\rm crit}$ for population II (Pop II) stars assuming idealized spectra of $\rm 10^{4}$ K was found to be about two orders of magnitude lower than Pop III stars. \cite{Agarwal12} performed N-body simulations along with semi-analytical models and found that Pop III are unable to produce the required critical flux for the DCBHs sites (see their Fig. 6) as they have very short lives and are less abundant. At the same time, trace amount of dust or metals are sufficient to produce Pop II stars \citep{2005ApJ...626..627O,2009A&A...496..365C,2013ApJ...766..103D,SafranekShrader14,BovinoCEMPs} which can more easily provide the flux to produce the observed abundance of quasars. As we show in this study using realistic Pop II spectra the critical value of the flux becomes comparable to that of Pop III stars and does not depend much on the stellar population. 

For realistic spectra of Pop II stars, $T_{\rm rad}$ is expected to be between $\rm 10^{4}-10^{5}$ K \citep{1999ApJS..123....3L,2003A&A...397..527S}. In a recent study, \cite{Omukai2014} have computed  Pop II spectra from the stellar synthesis code STARBURST \citep{1999ApJS..123....3L} and shown that the effect of realistic spectra can be mimicked by black body spectra with radiation temperatures between $\rm 10^{4}-10^{5}$ K. Their Fig. 6 demonstrates that the value of $J_{21}^{\rm crit}$ solely depends on the ratio of the H$^-$ to H$_2$ photo-dissociation rates and that black body radiation spectra with temperatures between $\rm 10^{4}-10^{5}$ K can reproduce the results of realistic Pop II spectra. The value of $J_{21}^{\rm crit}$ from their one-zone model varies from 1000-1400. They further found that $J_{21}^{\rm crit}$ does not depend on the age or metallicity for constant star formation and even decreases for instantaneous star formation in young metal poor galaxies, contrary to the findings of \cite{2014arXiv1407.4115A}.

Such actively star forming galaxies also produce X-rays as they host massive stars, X-ray binaries and possibly mini-quasars \citep{2005MNRAS.363.1069K,2014MNRAS.440.3778J,2014arXiv1404.2578S,2014arXiv1407.1847H}. X-rays have a long mean free path (small absorption cross-section for HI) and can easily escape to build up a cosmic X-ray background. They can photo-ionize/photo-heat the gas and may boost the formation of molecular hydrogen by enhancing the degree of ionization \citep{1997ApJ...476..458H,2003MNRAS.340..210G}. \cite{2011MNRAS.416.2748I} performed one-zone calculations to study the impact of the X-ray background flux in the context of the direct collapse scenario  and found that it raises the critical threshold for UV by a factor 5 ($J_{21}^{\rm crit} \propto J_{\rm X}^{1/2}$) by boosting $\rm H_{2}$ formation. 

In this study, we compute the value of $J_{21}^{\rm crit}$ for realistic Pop II spectra. To achieve this goal, we perform three-dimensional cosmological simulations for three distinct massive primordial halos of $\rm 10^7-10^8$~M$_{\odot}$ at $z>10$ by employing a comprehensive chemical model. We further compute the dependence of $J_{21}^{\rm crit}$ on the radiation spectra for $T_{\rm rad}$ between $10^4-10^5$ K, on variations from halo to halo, and on the impact of cosmic X-ray ionization. This is the first study exploring the impact of both UV and X-ray background radiation via three-dimensional cosmological simulations with a fixed Jeans resolution of 32 cells and employing a higher order chemical solver \citep[see][]{2013MNRAS.434L..36B}. We implement a comprehensive chemical model which takes into account all the chemical and thermal process relevant for this study. This work has significant implications for the expected number density of DCBHs as it provides constraints necessary for their formation.

The article is organized in the following way. In section 2, we provide the details of numerical methods and chemical model employed in this work. We present our findings in section 3 and confer our conclusions in section 4.

\section{Computational methods}
The simulations presented in this work are performed with the open source code ENZO\footnote{http://enzo-project.org/, changeset:48de94f882d8} \citep{2014ApJS..211...19B} which is an adaptive mesh refinement, grid based, parallel cosmological simulations code. The piece-wise parabolic solver is used to solve the hydrodynamics equations which is an improved form of the Godunov method  \citep{1984JCoPh..54..174C}. It makes use of the particle mesh technique to solve the dark matter (DM) dynamics and the multi-grid Poisson solver to compute the gravity.

Our simulations are commenced with cosmological initial conditions selected from Gaussian random fields at $z=100$. We first perform simulation with a uniform grid resolution of $\rm 128^3$ cells for hydrodynamics and $\rm 128^3$ dark matter particles to select the most massive halos at redshift 15 using the friend of friends algorithm \citep{2011ApJS..192....9T}. Our computational volume has a comoving size of 1 Mpc/h. The parameters from the WMAP seven years data \citep{2011ApJS..192...14J} are used for generating the initial conditions. We rerun the simulations by employing two nested refinement levels each with a grid resolution of $\rm 128^3$ in addition to the top grid resolution of $\rm 128^3$ both for the hydrodynamics and gravity. In total, we use 5767168 particles to solve the DM dynamics which yields an effective particle mass resolution of about 600 solar masses. We employ additional 18 refinement levels in the central 62 kpc region during the course of the simulations and resolve the Jeans length by 32 cells. Our refinement criteria are based on the gas overdensity and the particle mass resolution. The cells are marked for refinement if their density exceeds four times the cosmic mean density. Similarly, grid cells having DM densities above 0.0625 times  $ \rho_{DM}r^{\ell \alpha}$ are flagged for refinement where $r=2$ is the refinement factor, $\ell$ is the refinement level, and $\alpha =-0.3$ makes the refinement super-Lagrangian. We further smooth the dark matter particles at refinement level 12 which is equivalent to 2 pc in physical units to avoid spurious numerical artifacts. The simulations are stopped after reaching the maximum refinement level.

\subsection{Chemical model}

To follow the thermal and chemical evolution of the gas, we employ the publicly available package KROME\footnote{www.kromepackage.org, changeset:4674be5} and its built-in ENZO patch \citep{2014MNRAS.439.2386G}. The rate equations of $\rm H$, $\rm H^{+}$, $\rm He$, $\rm He^{+}$,~$\rm He^{++}$, $\rm e^{-}$,~$\rm H^{-}$,~$\rm H_{2}$,~$\rm H_{2}^{+}$ are solved self-consistently along with the hydrodynamics in cosmological simulations. The processes involving deuterium species are ignored as they do not affect the findings of this work. We further presume here that a uniform isotropic background UV flux of various intensities below the Lyman limit (i.e. below 13.6 eV) in units of $\rm 10^{-21}~erg~cm^{-2}~s^{-1}~Hz^{-1}~sr^{-1}$ is emitted by Pop II stars with black body radiation temperatures between $\rm 10^4-10^5$~K. As shown by \cite{Omukai2014}, such spectra effectively mimic the results from a realistic spectrum. In addition to this, we employ a cosmic X-ray background of various strengths (details are described below). Our chemical model is an extended form of \cite{2014MNRAS.443.1979L} with the addition of X-ray chemistry, updated rates and the $\rm H_{2}$ cooling function for low densities by \cite{2008MNRAS.388.1627G}. It further includes the photo-detachment of $\rm H^-$, the photo-dissociation of $\rm H_{2}$ and $\rm H_{2}^{+}$, collisional dissociation, collisional induced emission, chemical cooling/heating from three-body reactions, cooling by collisional excitation, collisional ionization, radiative recombination and bremsstrahlung radiation \citep{2014MNRAS.439.2386G}. For $\rm H_{2}$ cooling, we used the escape probability given in \cite{Omukai2000} which is based on the Sobolev approximation. The $\rm H_{2}$ self-shielding fitting function by \cite{2011MNRAS.418..838W} is employed here.

\begin{figure}
%  \vspace{-4.0cm}
 \hspace{-4.0cm}
\centering
\begin{tabular}{c}
\begin{minipage}{4cm}
\includegraphics[scale=0.6]{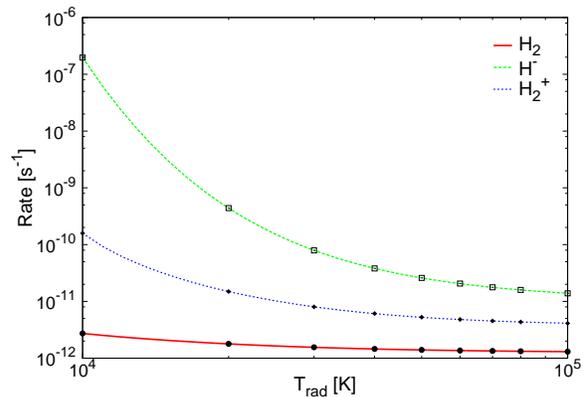}
\end{minipage}
\end{tabular}
\caption{Photo-dissociation rates of $\rm H_{2}$, $\rm H_{2}^{+}$ \& photo-detachment rate of $\rm H^-$ are plotted in this figure for various radiation temperatures between $T_{\rm rad}=10^4-10^5$ K and $J_{21}$=1. Different line styles represent the fitting functions provided in table \ref{table0} and the symbols show the original data. }
\label{fig0}
\end{figure}

% \subsubsection{X-rays}
Due to the smaller X-ray absorption cross-sections for hydrogen compared to the ionizing UV photons, X-rays can travel longer distances and build up a cosmic X-ray background (CXB). Following \cite{2003MNRAS.340..210G,2011MNRAS.416.2748I}, we assume that the CXB has a power law spectrum with index of -1.5 and is given by
\begin{equation}
J_{\rm X}(\nu) = J_{\rm X,21} \times 10^{-21} \left( \nu \over \nu_{0} \right)^{-1.5}~ \mathrm{ erg~cm^{-2}~s^{-1}~Hz^{-1}~sr^{-1}}
\label{eqx}
\end{equation}
where $h \nu_{0}$=1 keV and $ J_{\rm X,21}$ is CXB flux in units of $\rm 10^{-21}~erg~cm^{-2}~s^{-1}~Hz^{-1}~sr^{-1}$. We also consider both primary and secondary ionization of H and He atoms. We ignore the contribution from the secondary ionization of HeII which is negligible in our case as also found in the previous studies \citep{Shull1985}. For the cosmic X-ray background, we consider photons between 2-10 keV as low energy X-ray photons are absorbed locally \citep{1997ApJ...476..458H}. Further details about the implementation of X-ray physics, photo-chemistry and reactions rates are provided in the appendix.
% \begin{equation}
% \varGamma_{X} = \varGamma_{X,H}+ \varGamma_{X,He}
% \end{equation}
% \begin{equation}
% \Gamma_{X,i} = \int {4\pi J_{X}(\nu) \over h\nu} e^{-\tau_{\nu}} \sigma_{i}(\nu)E_{h,i} d\nu
% \end{equation}
% where i=H, He and $\tau_{\nu}=N_{H}\sigma_{H}(\nu)+ N_{He}\sigma_{He}$, E is the energy deposited as heat by ionization and is given by \cite{1995ApJ...443..152W}. 
% 

\begin{table*}
\begin{center}
\caption{ The fitting functions for the photo-dissociation of $\rm H_{2}$, $\rm H_{2}^{+}$ and the photo-detachment of $\rm H^{-}$. They are valid for $T_{\rm rad}=10^4-10^5$ K.}
\begin{tabular}{clllllll}
\hline
\hline

 Coefficients	& $ k_{\rm H_{2}}[s^{-1}]$ & $k_{\rm H^{-}}[s^{-1}]$  & $ k_{\rm H_{2}^{+}}[s^{-1}]$ \\
\hline
& dex$[(a + bT_{\rm rad} + cT_{\rm rad}^2)^{-1} - d]$  & dex$[(a + bT_{\rm rad})^{-1/c} - d] $ &  $(-a + bT_{\rm rad})^{-1/c} + d$  \\
\hline                                                          
 
% $\rm 1 \times  10^{4}$ & $\rm 1.9666 \times 10^{-7}$     & $\rm 2.7283 \times 10^{-12}$    &  $\rm 1.5883  \times 10^{-10}$ \\
% $\rm 2 \times 10^{4}$  & $\rm 4.3945  \times 10^{-10}$   & $\rm 1.7866  \times 10^{-12}$   &  $\rm 1.4937 \times 10^{-11}$ \\
% $\rm 3 \times 10^{4}$  & $\rm 7.9028  \times 10^{-11}$   & $\rm 1.55372  \times 10^{-12}$  &  $\rm 8.0044 \times 10^{-12}$ \\
% $\rm 4 \times 10^{4}$  & $\rm 3.8012 \times  10^{-11}$   & $\rm 1.45210 \times  10^{-12}$  &  $\rm 6.1006 \times 10^{-12}$ \\
% $\rm 5 \times 10^{4}$  & $\rm 2.5944 \times  10^{-11}$   & $\rm 1.39688 \times  10^{-12}$  &  $\rm 5.2672 \times 10^{-12}$ \\
% $\rm 6 \times 10^{4}$  & $\rm 2.0605 \times  10^{-11}$   & $\rm 1.36288 \times 10^{-12}$   &  $\rm 4.8130 \times 10^{-12}$ \\
% $\rm 7 \times 10^{4}$  & $\rm 1.7709  \times 10^{-11}$   & $\rm 1.34015 \times  10^{-12}$  &  $\rm 4.5316 \times 10^{-12}$ \\
% $\rm 8 \times 10^{4}$  & $\rm 1.5923 \times 10^{-11}$    & $\rm 1.32401 \times  10^{-12}$  &  $\rm 4.3420 \times 10^{-12}$ \\
% $\rm 1 \times 10^{5}$  & $\rm 1.3867 \times 10^{-11}$    & $\rm 1.30282 \times 10^{-12}$   &  $\rm 4.1047  \times 10^{-12}$ \\
% H2
a  & $\rm 1.1735 \times 10^{-1}$    & $\rm 9.08944 \times 10^{-2}$   &  $\rm 3.83012  \times 10^{6}$ \\
b  & $\rm  2.4958\times 10^{-4}$    & $\rm  3.27940 \times 10^{-5}$ &  $\rm  5.06440  \times 10^{2}$ \\
c  & $\rm  3.4856 \times 10^{-9}$    & $\rm 5.98490 \times 10^{-1}$   &  $\rm  6.20988 \times 10^{-1}$ \\
d  & $\rm 1.1902 \times 10^{1}$    & $\rm 1.09867 \times 10^{1}$   &  $\rm 3.68778 \times 10^{-12}$ \\

% k(H2) = dex[(a + b*Trad + c*Trad^2)^-1 + d]
% 
% 
% HM
% 
% k(HM) = dex[(a + b*Trad)^(-1/c) + d]
% 
% H2+
% 
% k(H2+) = (a + b*Trad)^(-1/c) + d

\hline
\end{tabular}
\label{table0}
\end{center}
\end{table*}

\begin{figure*}
%  \vspace{-4.0cm}
\hspace{-6.0cm}
\centering
\begin{tabular}{c}
\begin{minipage}{6cm}
\includegraphics[scale=1.0]{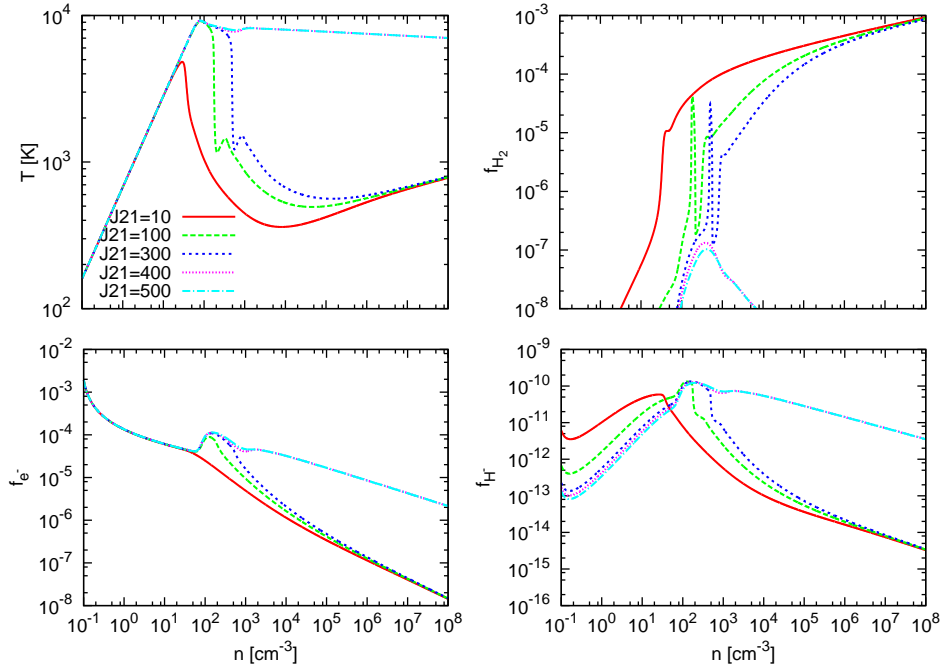}
\end{minipage}
\end{tabular}
\caption{One-zone plot for $T_{\rm rad}=2 \times 10^4$ K. The temperature and the abundances of $\rm H_{2}$, $\rm e^{-}$ and $\rm H^{-}$ are plotted against the number density for various strengths of a background UV flux.}
\label{fig1}
\end{figure*}

\begin{figure*}
%  \vspace{-4.0cm}
\hspace{-6.0cm}
\centering
\begin{tabular}{c}
\begin{minipage}{6cm}
\includegraphics[scale=1.0]{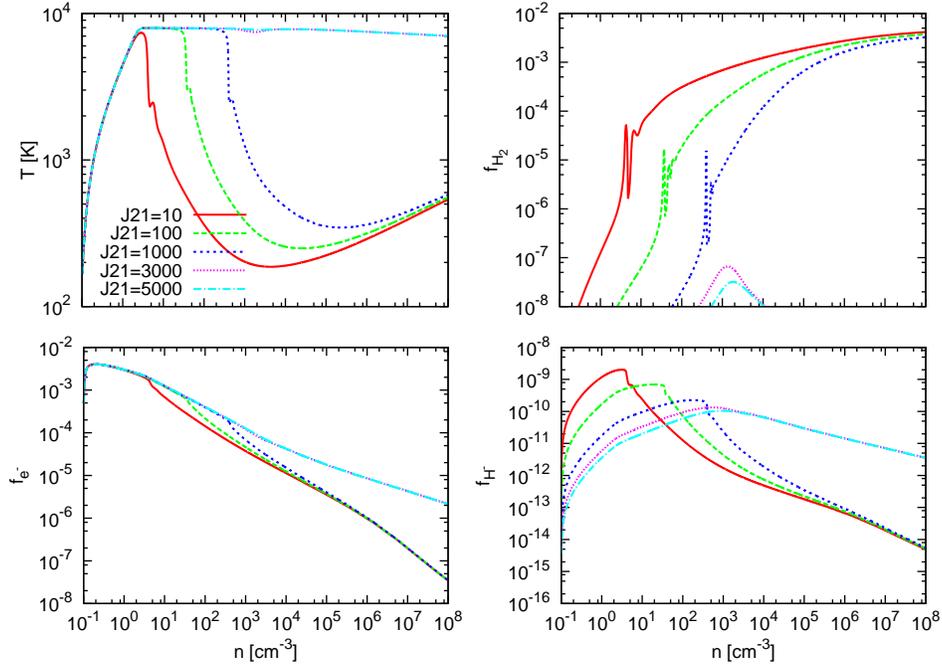}
\end{minipage}
\end{tabular}
\caption{ Same as figure \ref{fig1} for the X-ray flux of $J_{\rm X,21}=0.1$.}
\label{fig2}
\end{figure*}

\begin{figure*}
\centering
% \hspace{9cm}
\begin{tabular}{c c}
\begin{minipage}{4cm}
\hspace{-5cm}
\includegraphics[scale=0.45]{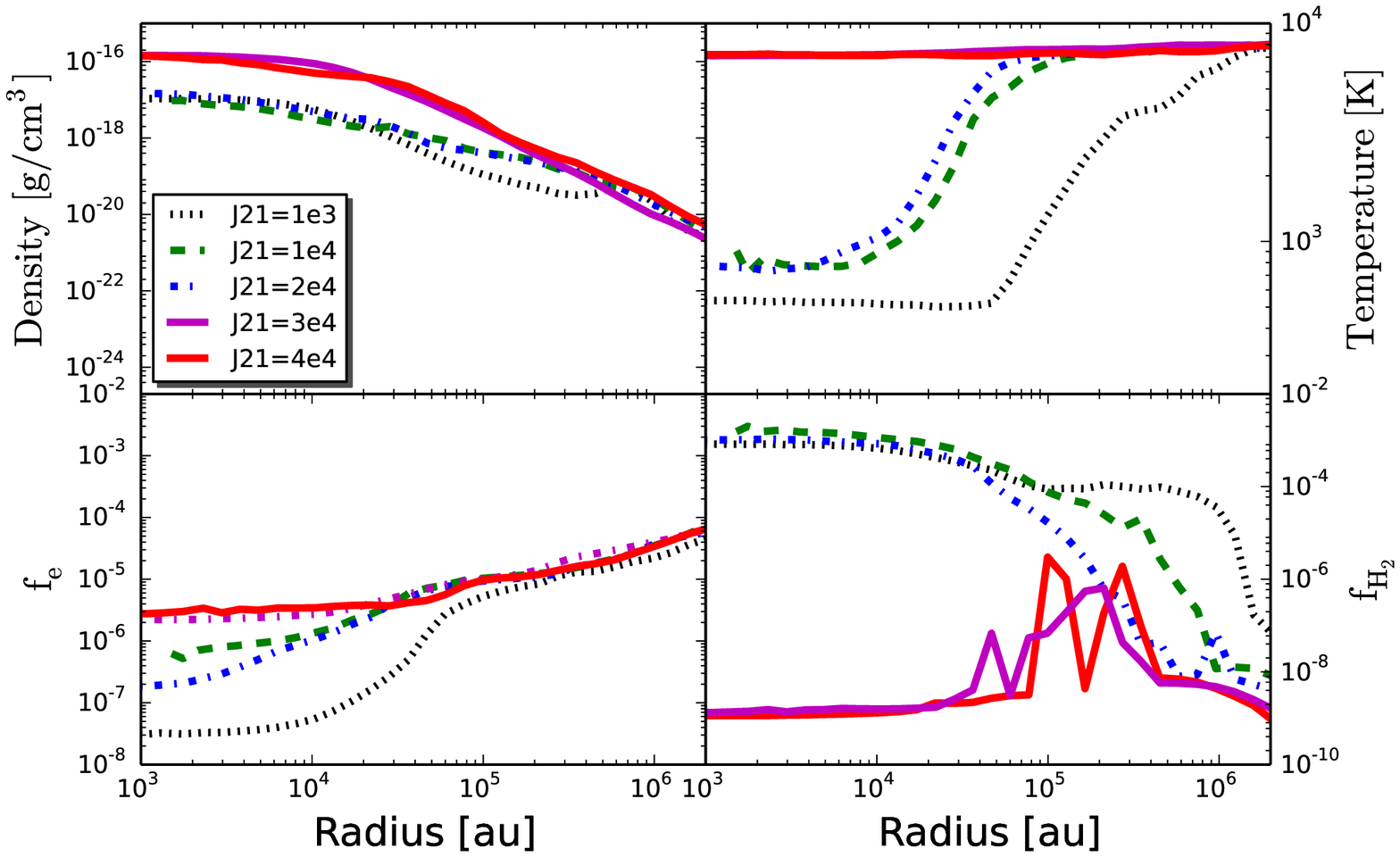}
\end{minipage} &
\begin{minipage}{4cm}
% \hspace{1cm}
\includegraphics[scale=0.45]{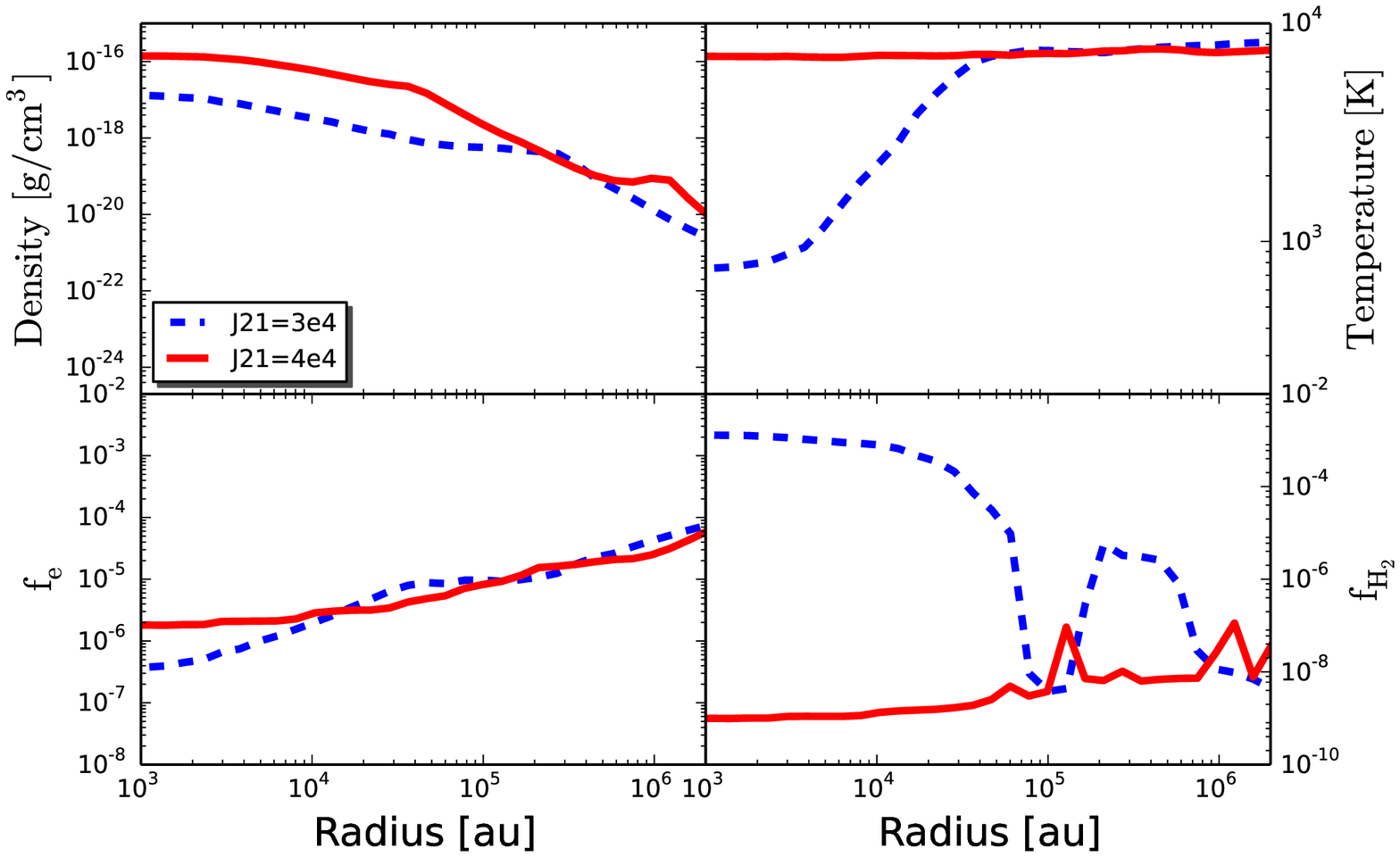}
\end{minipage} \\
\begin{minipage}{4cm}
\includegraphics[scale=0.45]{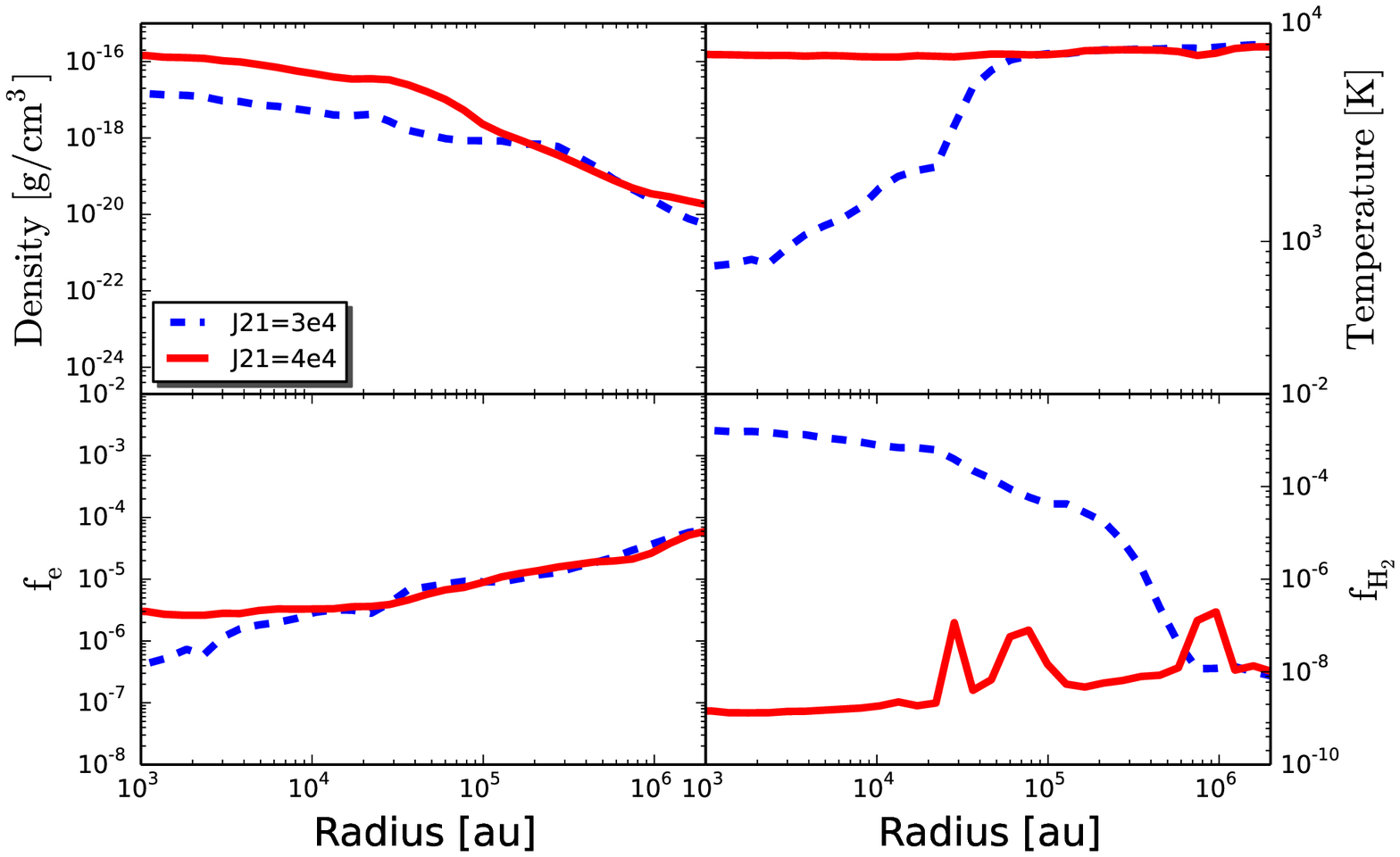}
\end{minipage}
\end{tabular}

\caption{Spherically averaged and radially binned profiles of temperature, density, $\rm H_{2}$ and $\rm e^{-}$ fractions computed for the peak density in the simulations are plotted for the halo A. The top panels represent $T_{\rm rad}=2.0 \times 10^4$ K (left) $T_{\rm rad}=4.0 \times 10^4$ K (right) and the bottom panel $T_{\rm rad}=8.0 \times 10^4$ K. Each line style depicts the value of $J_{21}$ as mentioned in the legend. }
\label{fig3}
\end{figure*}

\begin{figure*}
%  \vspace{-4.0cm}
% \hspace{-6.0cm}
\centering
\begin{tabular}{c c}
\begin{minipage}{4cm}
\hspace{-5cm}
\includegraphics[scale=0.45]{Trad2e4FirstHalo.eps}
\end{minipage} &
\begin{minipage}{4cm}
% \hspace{1cm}
\includegraphics[scale=0.45]{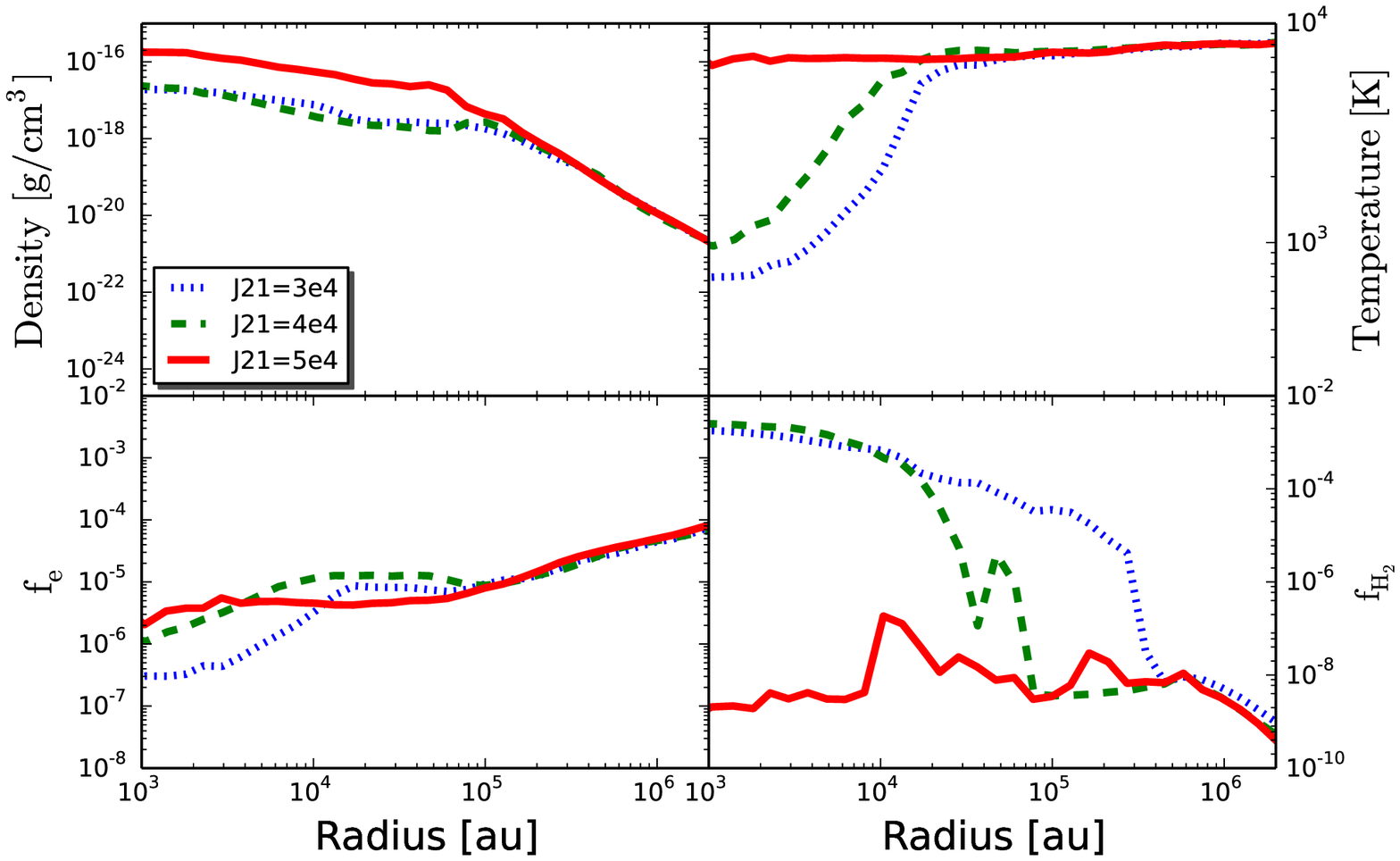}
\end{minipage} \\
\begin{minipage}{4cm}
\includegraphics[scale=0.45]{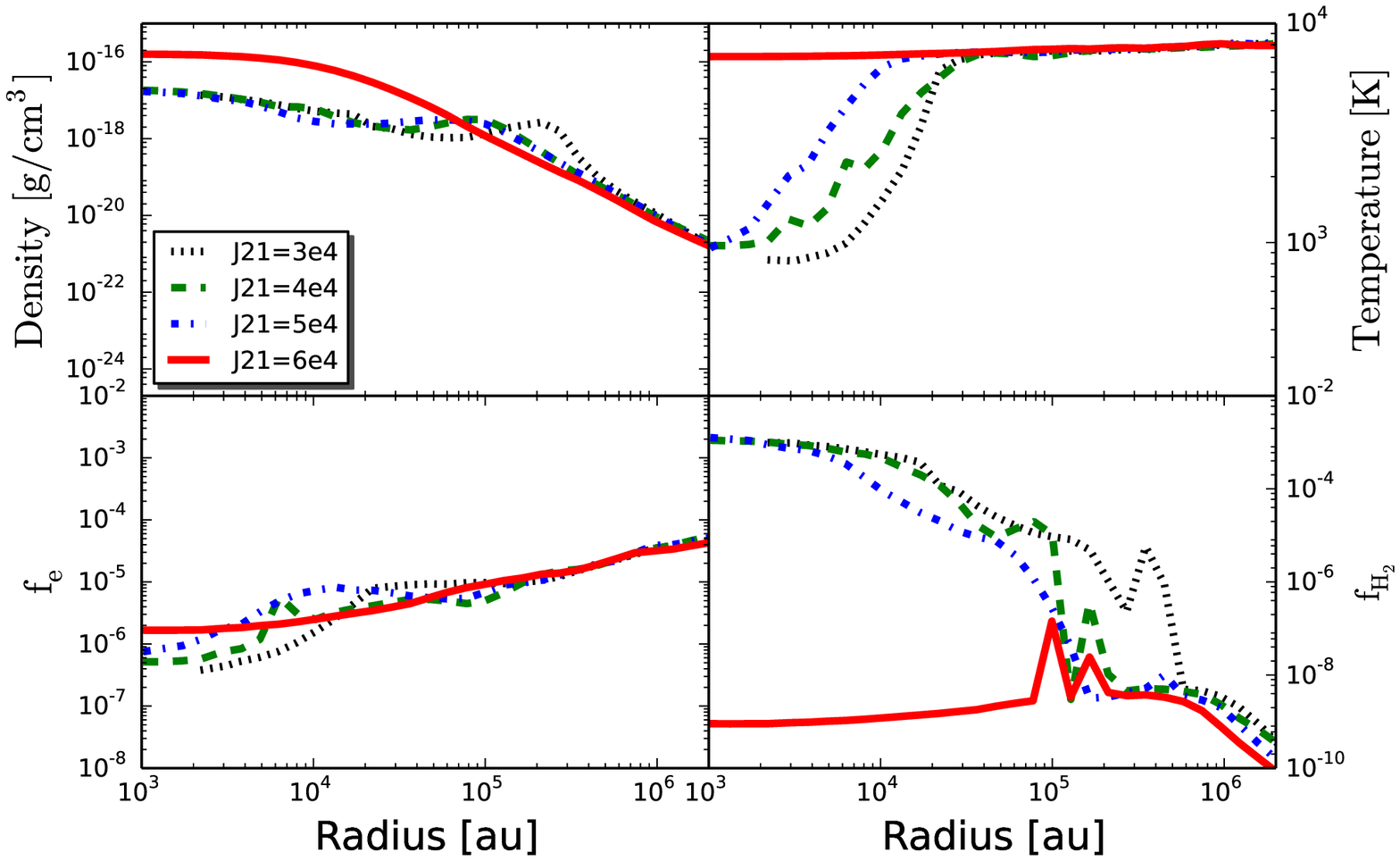}
\end{minipage}
\end{tabular}
\caption{Spherically averaged and radially binned profiles of temperature, density, $\rm H_{2}$ and $\rm e^{-}$ fractions computed for the peak density in the simulations are plotted for $T_{\rm rad}=2.0 \times 10^4$ K. The top panels represent halos A \& B (left to right) and the bottom panel halo C. Each line style depicts the value of $J_{21}$ as mentioned in the legend. The collapse redshifts of these halos are listed in table \ref{table1}}.
\label{fig4}
\end{figure*}

\begin{figure*}
\centering
\begin{tabular}{c c}
\begin{minipage}{4cm}
\hspace{-5cm}
\includegraphics[scale=0.45]{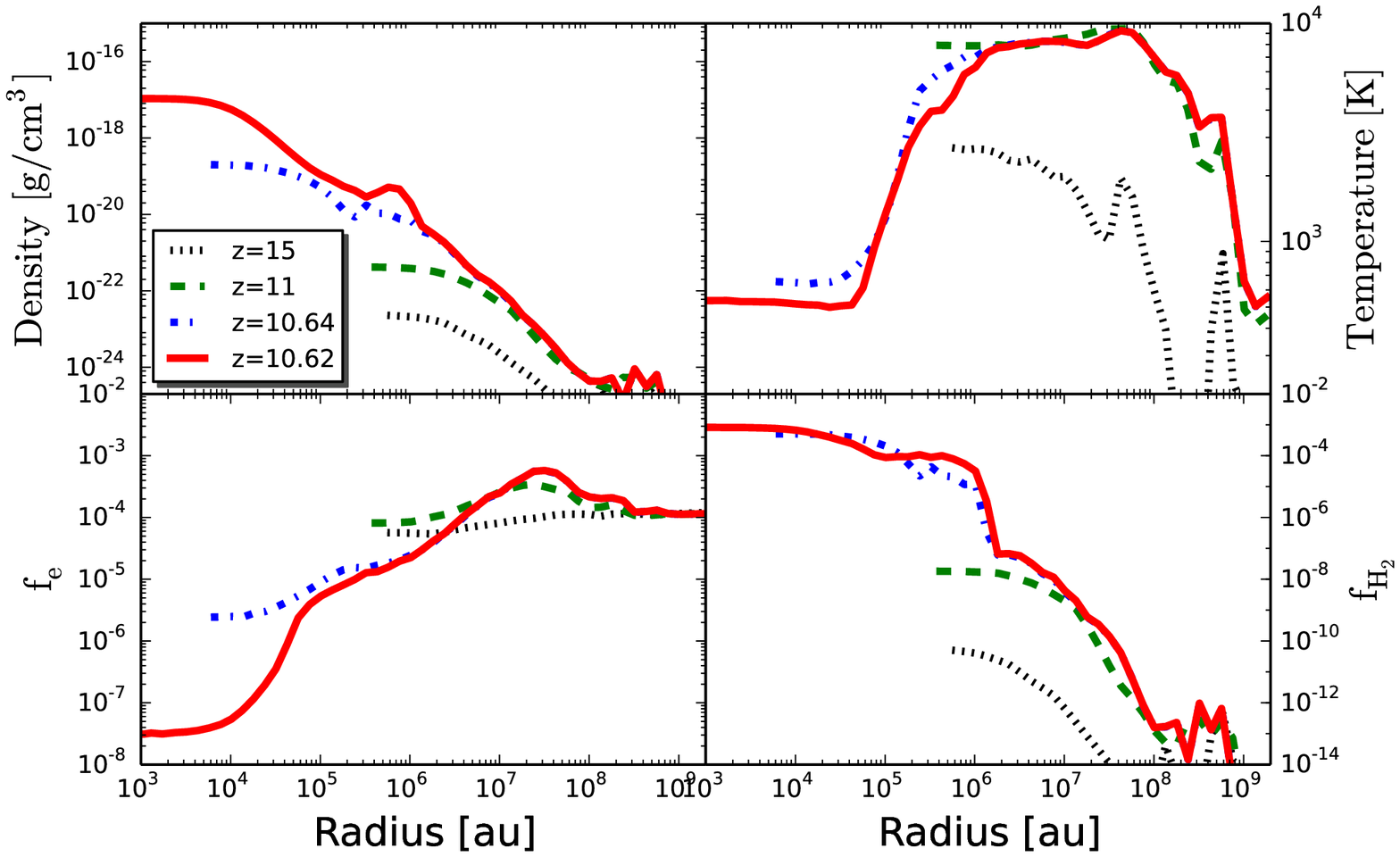}
\end{minipage} &
\begin{minipage}{4cm}
% \hspace{1cm}
\includegraphics[scale=0.45]{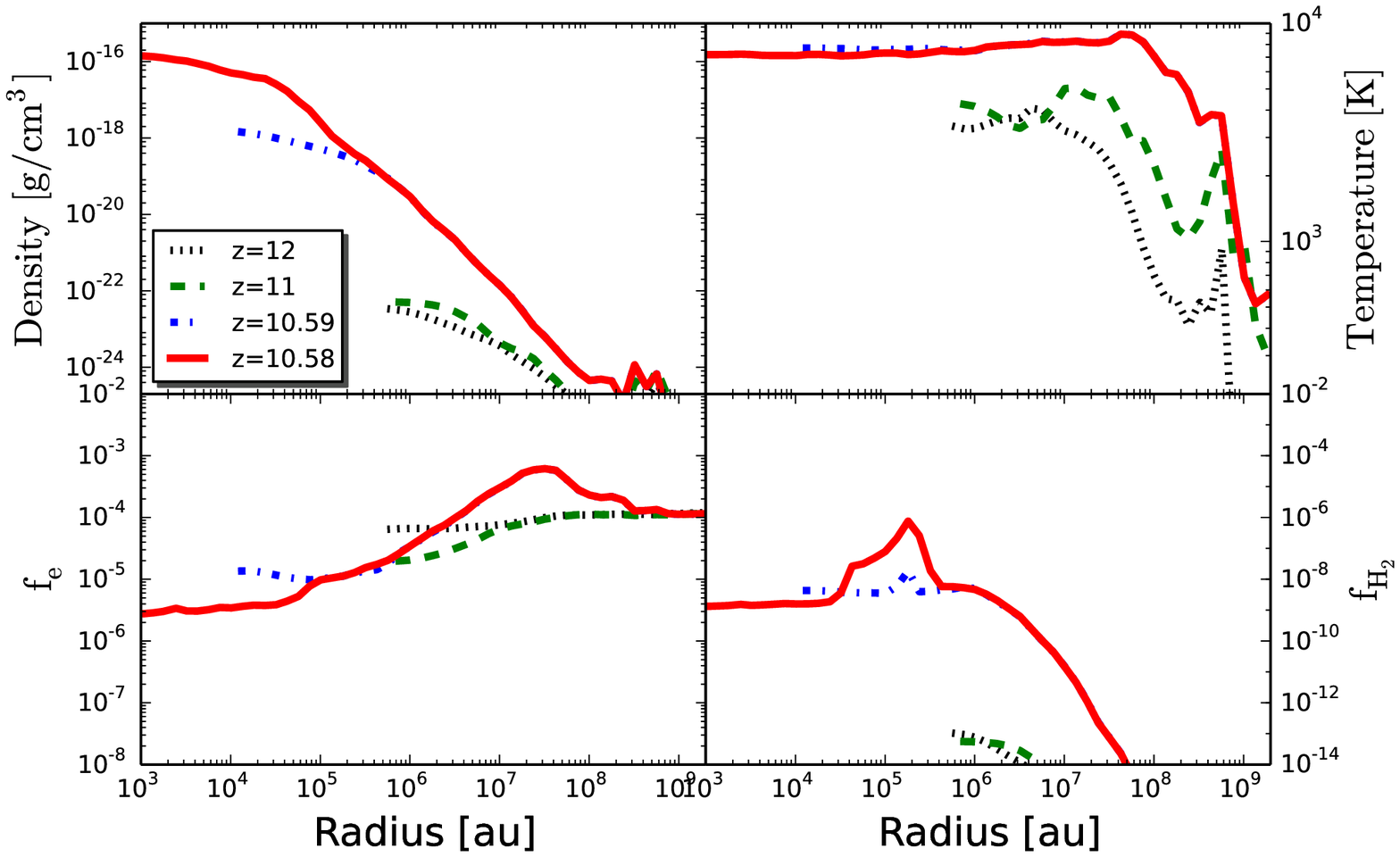}
\end{minipage} 
\end{tabular}
\caption{Time evolution of spherically averaged and radially binned profiles of temperature, density, $\rm H_{2}$ and $\rm e^{-}$ fractions is shown for the halo A and $T_{\rm rad}=2.0 \times 10^4$ K. Each line style represents a different redshift as shown in the legend. The left panel shows the $\rm H_2$ cooling case and the right panel depicts atomic cooling case.}
\label{fig41}
\end{figure*}

\begin{figure*}
%  \vspace{-4.0cm}
% \hspace{-6.0cm}
\centering
\begin{tabular}{c c}
\begin{minipage}{4cm}
\hspace{-5cm}
\includegraphics[scale=0.45]{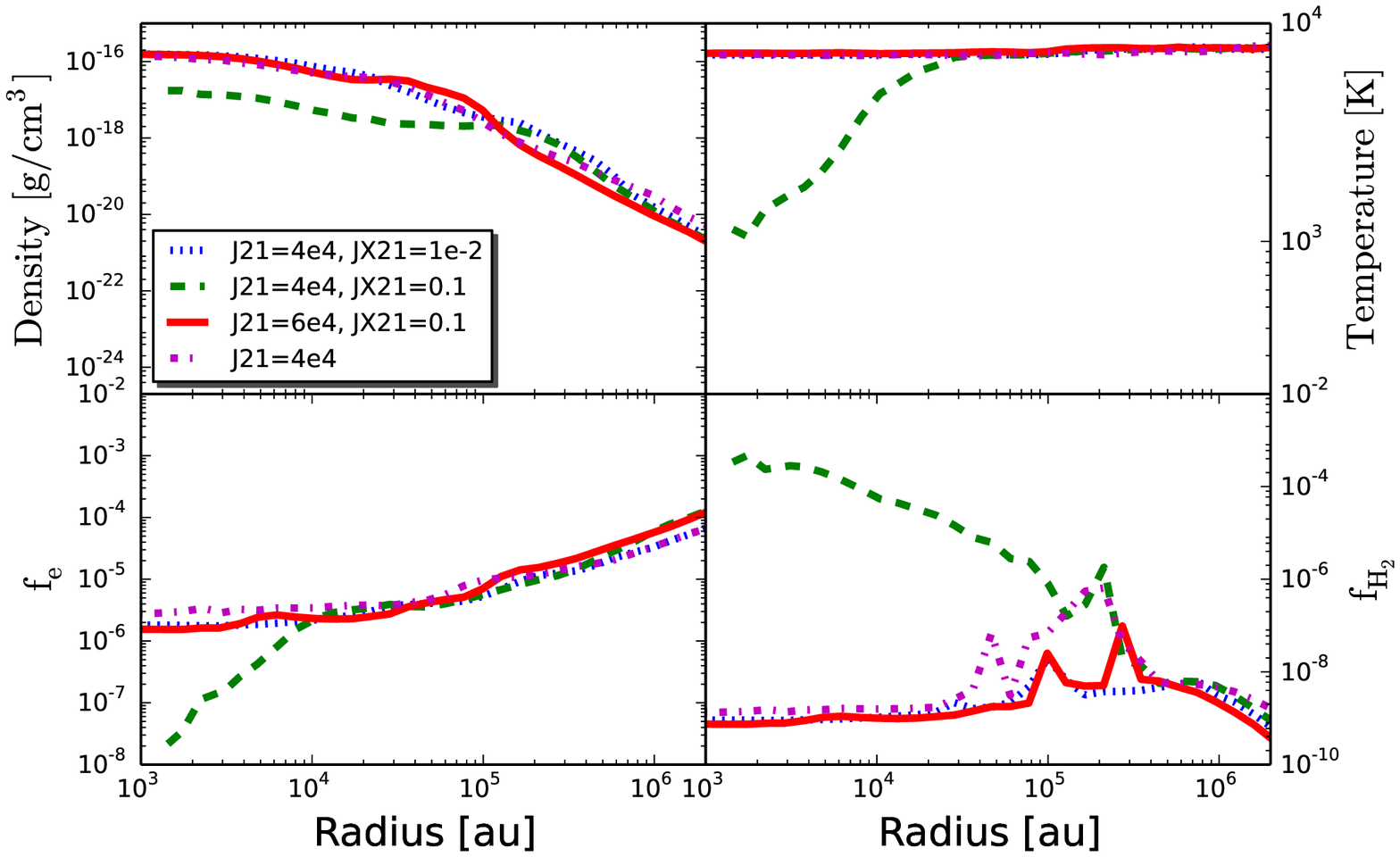}
\end{minipage} &
\begin{minipage}{4cm}
% \hspace{1cm}
\includegraphics[scale=0.45]{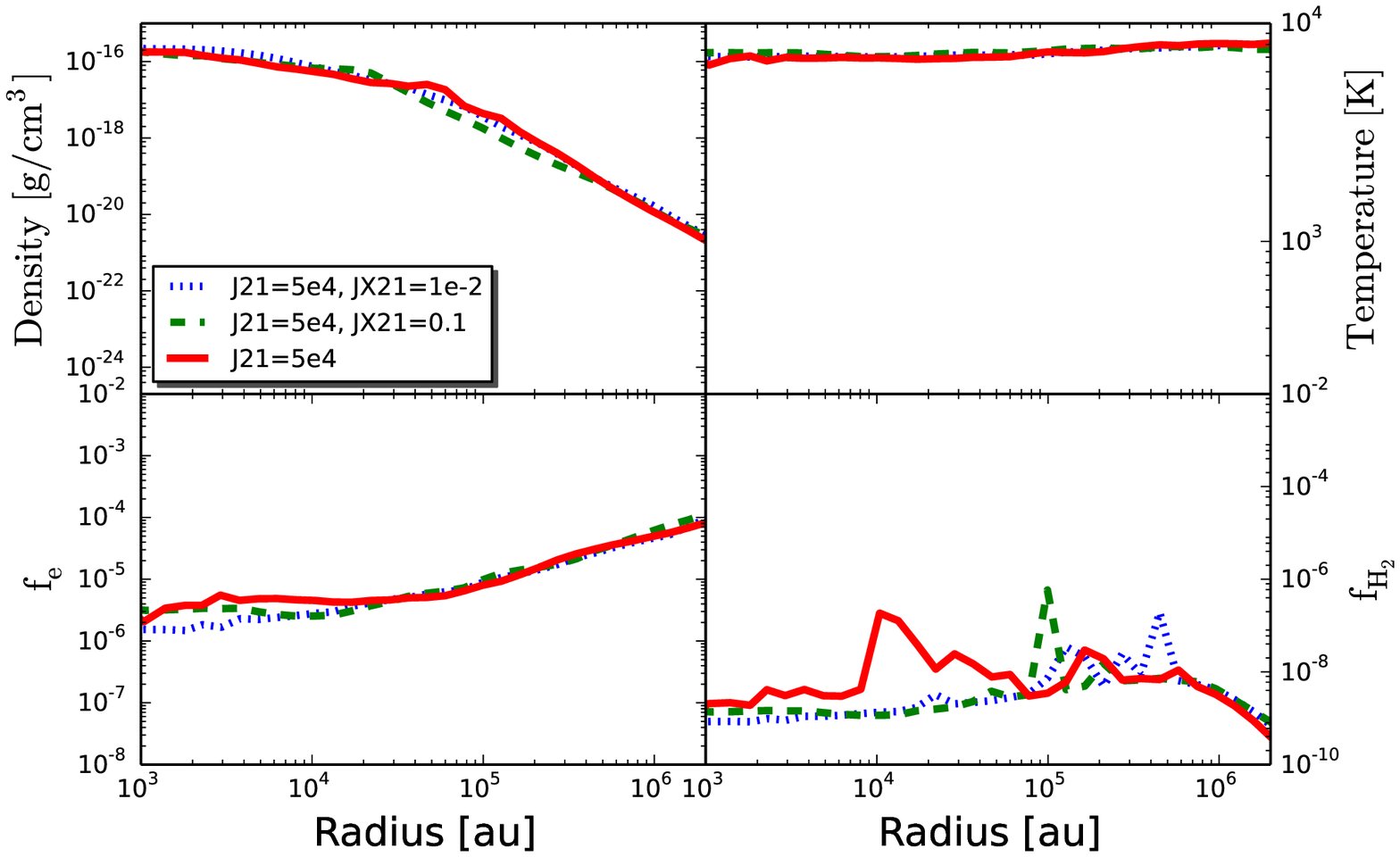}
\end{minipage} \\
\begin{minipage}{4cm}
\hspace{-5cm}
\includegraphics[scale=0.45]{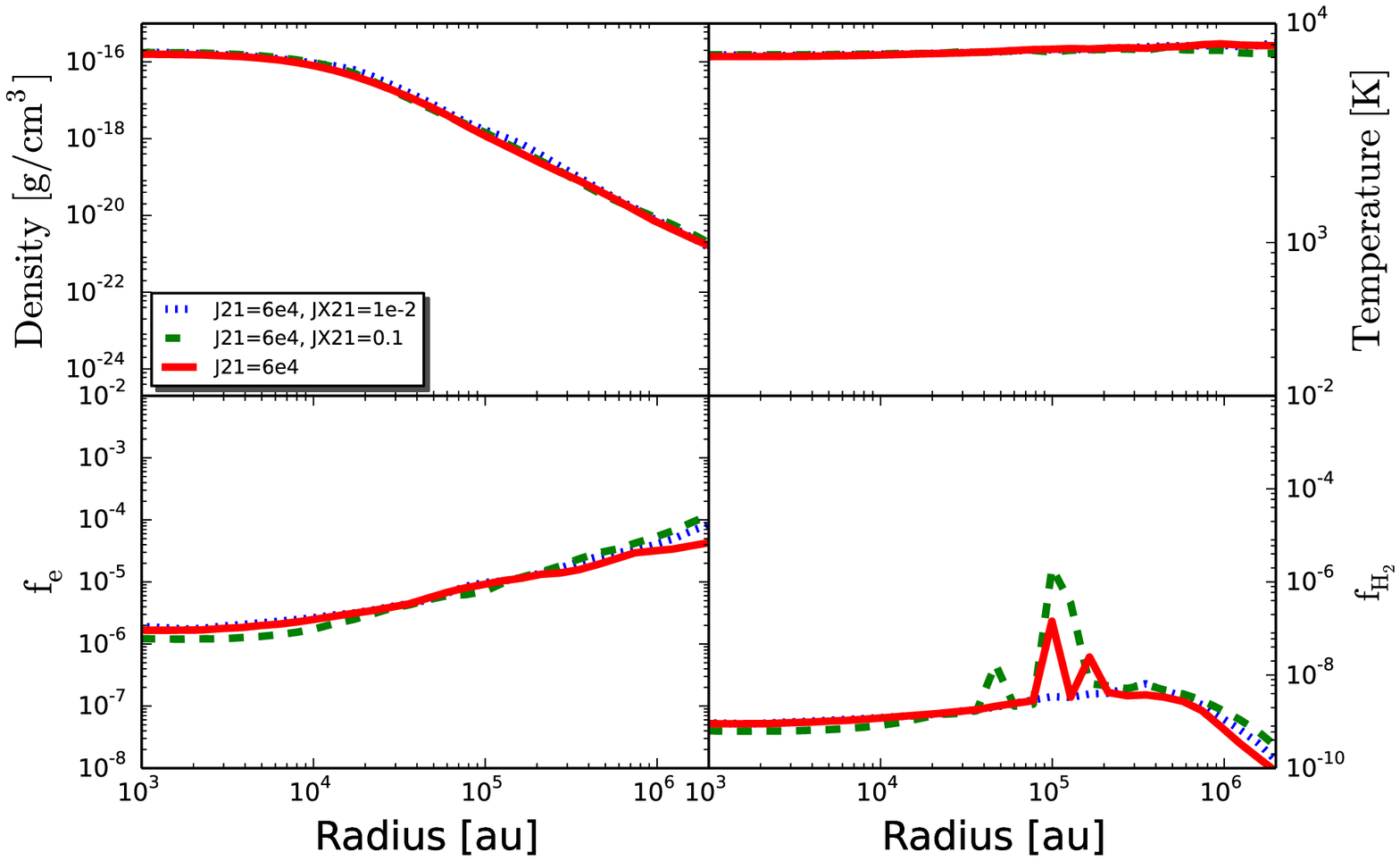}
\end{minipage} & 
\begin{minipage}{4cm} 
\includegraphics[scale=0.45]{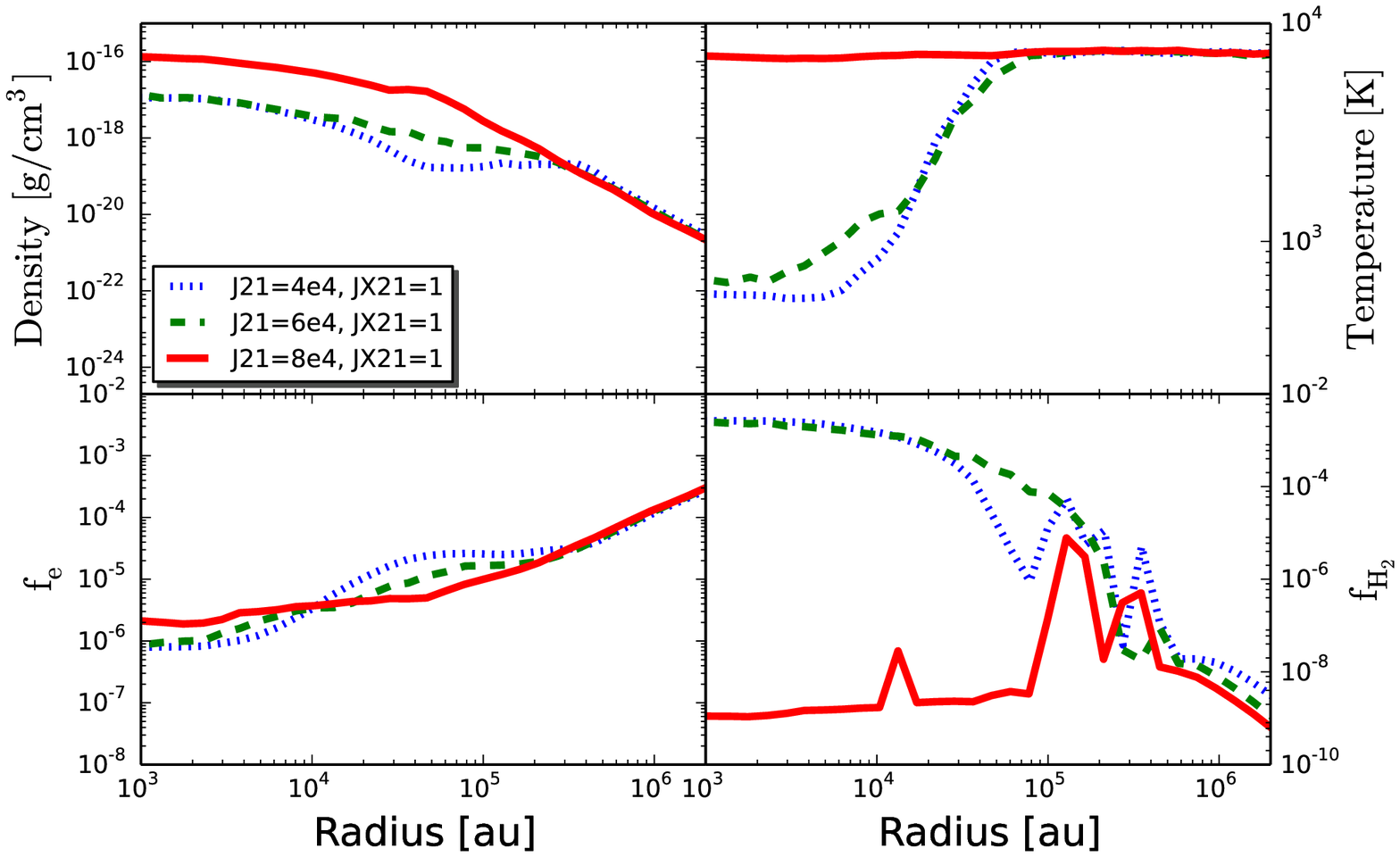}
\end{minipage}
\end{tabular}
\caption{Spherically averaged and radially binned profiles of temperature, density, $\rm H_{2}$ and $\rm e^{-}$ fractions computed for the peak density in the simulations are plotted for $T_{\rm rad}=2.0 \times 10^4$ K and {also include X-ray background.} The top panels represent halos A \& B (left to right) and the bottom left panel halo C while the right panel for halo A with higher X-ray flux. Each line style depicts the value of $J_{21}$ as mentioned in the legend.  }
\label{fig5}
\end{figure*}

\begin{figure*}
\centering
\begin{tabular}{c c}
\begin{minipage}{4cm}
\hspace{-5cm}
\includegraphics[scale=0.4]{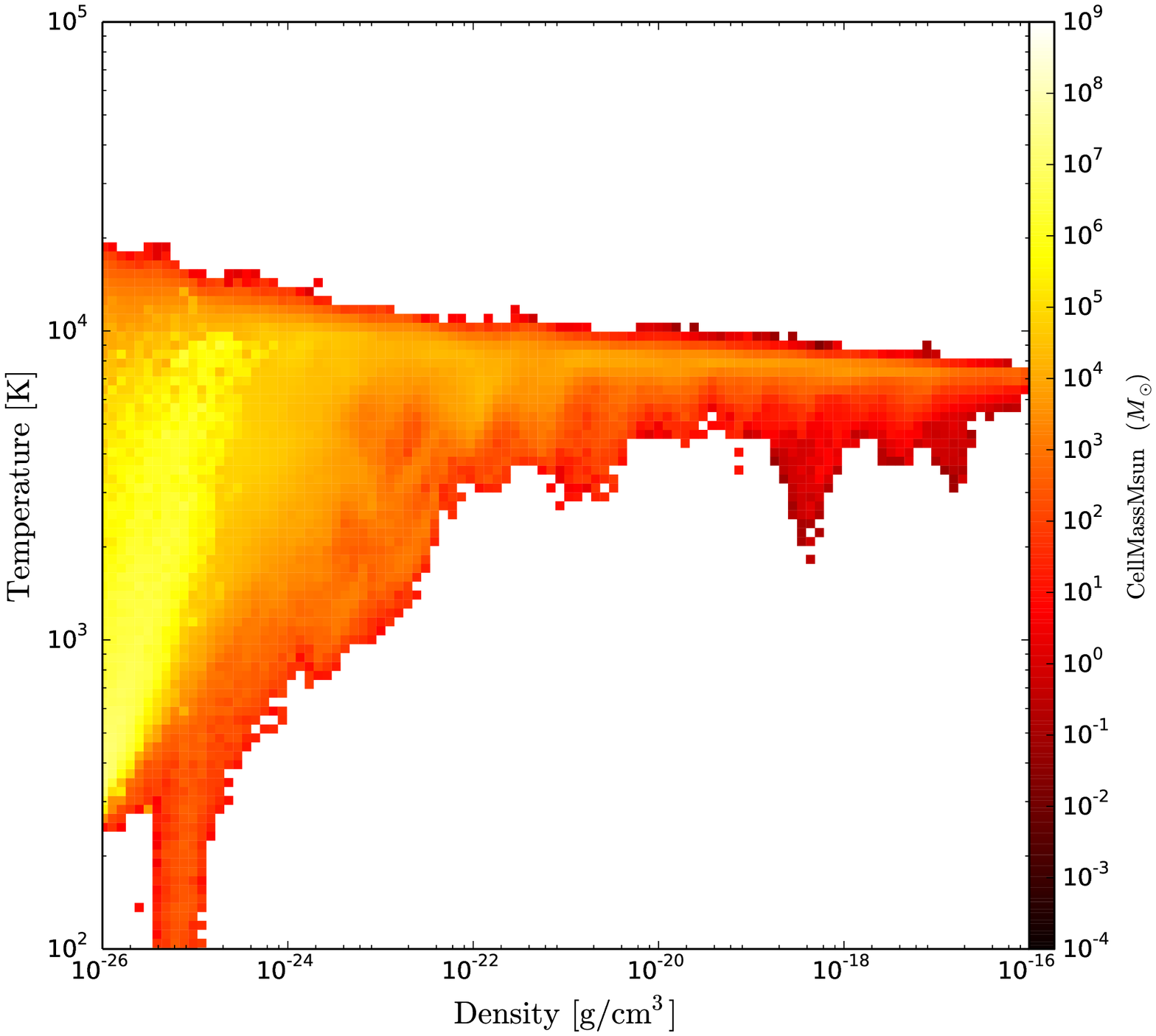}
\end{minipage} &
\begin{minipage}{4cm}
% \hspace{1cm}
\includegraphics[scale=0.4]{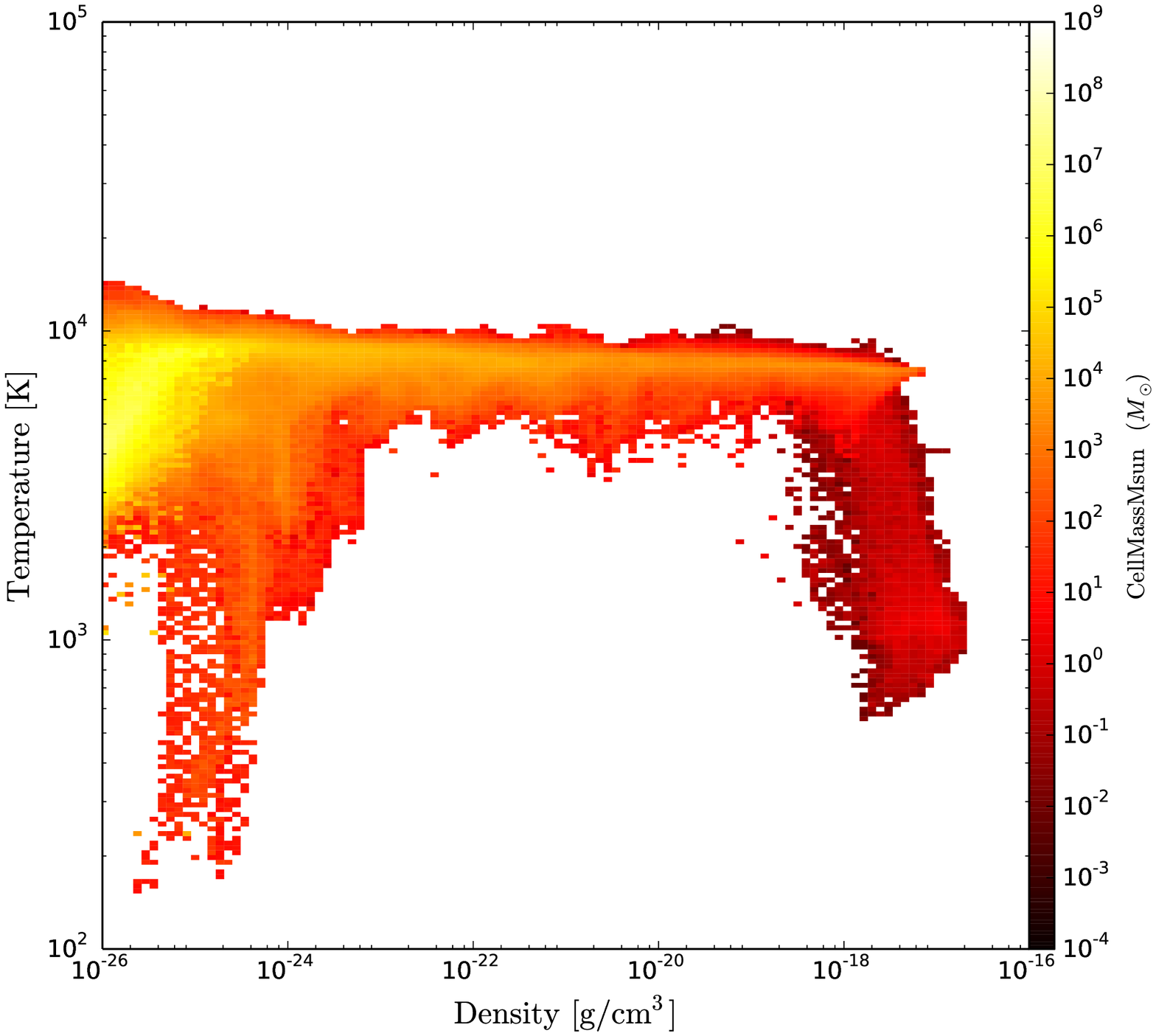}
\end{minipage} \\
\begin{minipage}{4cm}
\hspace{-5cm}
\includegraphics[scale=0.4]{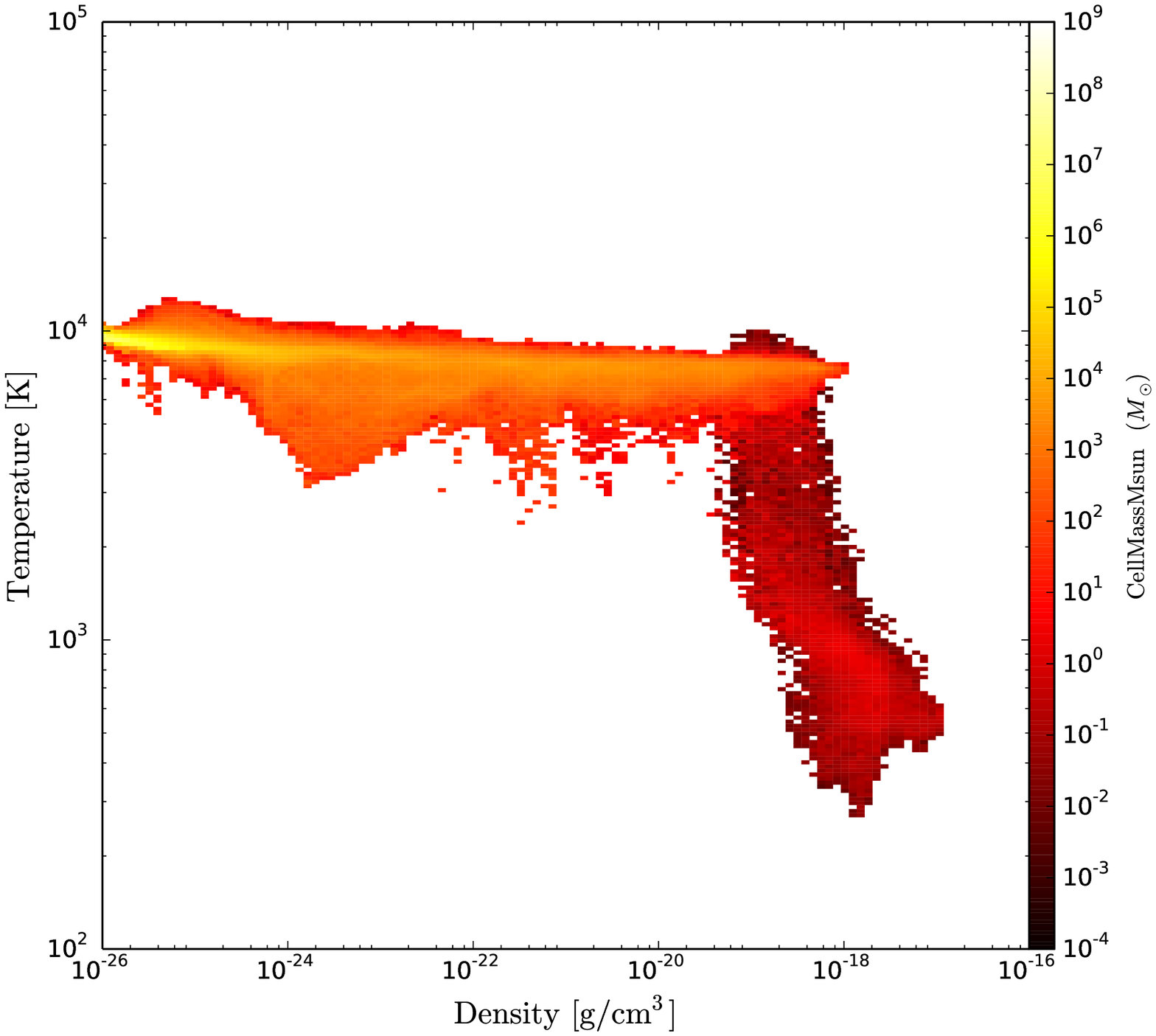} 
\end{minipage} &
\begin{minipage}{4cm}
\includegraphics[scale=0.4]{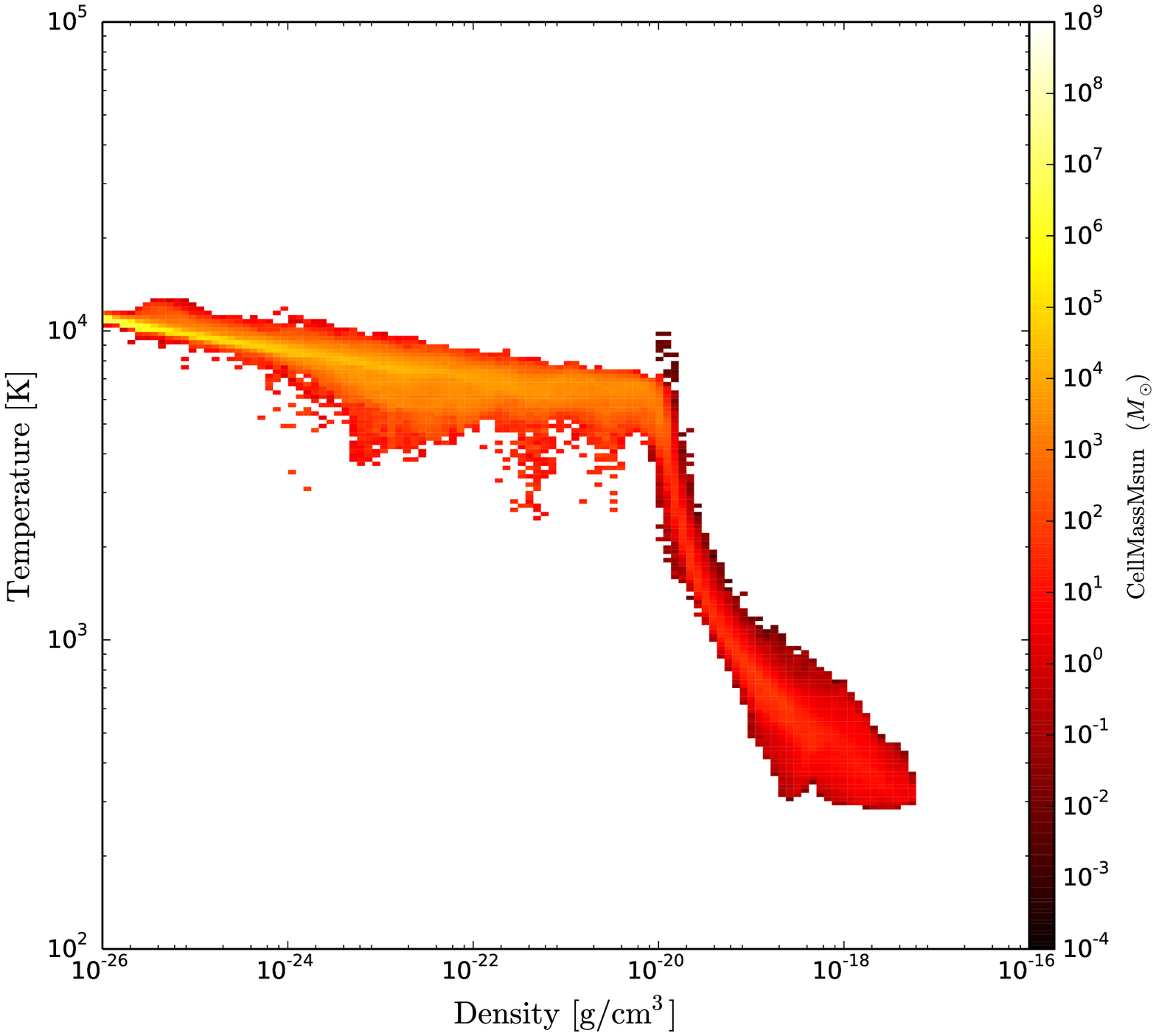} 
\end{minipage} 
\end{tabular}
\caption{ The phase plots (temperature verses density) of the halo A for $T_{\rm rad}=2 \times 10^4$ K and various value of $J_{\rm X,21}$. The top two panels represent $J_{\rm X,21}=0.01$ \&  $J_{\rm X,21}=0.1$ while the bottom two panels represent $J_{\rm X,21}=1$ \&  $J_{\rm X,21}=10$ from left to right, respectively.} 
\label{fig51}
\end{figure*}

\begin{table*}
\begin{center}
\caption{Properties of the simulated halos for $J_{21}^{\rm crit}$ are listed here.}
\begin{tabular}{ccccccc}
\hline
\hline

Model	& Mass	 & Redshift  & $J_{21}^{\rm crit}$ & $J_{\rm X,21}^{crit}$ & spin parameter\\

No & $\rm M_{\odot} $   & $z$  &  $T_{\rm rad}=2 \times 10^4$  K & in units of $J_{\rm X,21}$ & $\lambda$ \\
\hline                                                          \\
 		
A     & $\rm 5.6 \times 10^{7}$   &10.59  &20000 &40000 &0.034\\
B     & $\rm 4.06 \times 10^{7}$  &13.23  &40000 &40000 &0.02\\
C     & $\rm 3.25 \times 10^{7}$  &11.13  &50000 &50000 &0.03\\

\hline
\end{tabular}
\label{table1}
\end{center}
\end{table*}

\section{Main Results}

\subsection{Results from one-zone models}

We have computed the photo-dissociation rates of $\rm H_{2}$, $\rm H_{2}^{+}$ as well as the photo-detachment of $\rm H^{-}$ which depend on the adapted radiation spectra and play a key role in determining the $J_{21}^{\rm crit}$. These rates for $T_{\rm rad}=10^4-10^5$ K are not available in the literature. Therefore, we provide the fitting functions which are listed in table \ref{table0} and also shown in figure \ref{fig0}, see the appendix for detailed derivations (also see Fig. 4 in \cite{Omukai08}). It is noted that the $\rm H^-$ photo-detachment reaction varies by about 3 orders of magnitude for $T_{\rm rad}=10^4-2 \times 10^4$ K and its dependence becomes very weak between $T_{\rm rad}= 3 \times 10^4 - 10^5$ K. The $\rm H_{2}$  photo-dissociation reaction varies by a factor of a few between $T_{\rm rad}=10^4-10^5$ K. The same is the case for $\rm H_{2}^{+}$ photo-dissociation reaction.
% The reaction rates are computed by integerating the the cross-section for a given radiation spectra as explained in the appendix,

To verify our chemical model described in the above section, we have performed a number of one-zone tests and show some representative cases here. For the one-zone test, we took the initial gas temperature of 160 K, an initial gas density of $\rm 0.1~cm^{-3}$ and initial abundances of the species of e$^- = 2 \times 10^{-4}$ and H$_{2}= 2 \times 10^{-6}$, same as in \cite{2011MNRAS.416.2748I}. Figure \ref{fig1} depicts the thermal evolution and the abundances of $\rm H^{-}$, $\rm H_{2}$ and $\rm e^{-}$ for $T_{\rm rad}= 2 \times 10^4$ K. It is found that with increasing strength of the impinging radiation field the thermal evolution splits into two distinct tracks, the $\rm H_{2}$ and the atomic line cooling tracks. This trend is also evident in the abundance of $\rm H_{2}$ where in the former case it reaches the universal abundance (i.e. $\rm 10^{-3}$) while for the latter case, it remains a few orders of magnitude lower. This bifurcation defines the critical value of the incident flux which is found to be $J_{21}^{\rm crit}=300$ for a radiation spectrum of $\rm 2 \times 10^4$ K. This is an order of magnitude higher compared to the case of $T_{\rm rad}= 10^4$ K.  For the higher radiation temperature, the value of $J_{21}^{\rm crit}$ does not increase much as evident from the reaction rates discussed above and also clear from figure \ref{fig6}.

In comparison with previous studies, our results in general are consistent with \cite{Shang10} and \cite{Omukai2014}. However, the value of $J_{21}^{\rm crit}$ is a factor of a few lower from \cite{Omukai2014}. This difference mainly arises by considering the effect of dissociative tunneling for the collisional dissociation of $\rm H_{2}$ \citep{1996ApJ...461..265M}. It reduces the value of $J_{21}^{\rm crit}$ by a factor of 3. A similar effect was also observed by \cite{2014MNRAS.443.1979L}.

Motivated by the work of \cite{2011MNRAS.416.2748I} who found that the presence of a strong CXB further elevates the value of $J_{21}^{\rm crit}$, we have performed one-zone calculations including the X-ray background feedback. In figure \ref{fig2}, we show the impact of X-ray ionization for a CXB strength of $J_{\rm X,21}=0.1$ and $T_{\rm rad}= 2 \times 10^4$ K. It is found that for $J_{\rm X,21} < 0.01$ the impact of X-rays is negligible and starts to become important around $J_{\rm X,21} = 0.1$ where gas is instantly heated to 8000 K, the cooling threshold for the Lyman alpha line. A similar effect of X-ray heating has been observed in one-zone calculations as well as in 3D simulations \citep{Wolfire1995,Inayoshi2011,2014arXiv1407.1847H}. It has been found that X-ray heating is only important at low densities  $\rm < 1~cm^{-3}$ and the heating rate is of the order of $\rm 10^{-4}~erg~s^{-1}~g^{-1}$ which is a factor of 10 higher than other heating sources such as compressional heating. For a given intensity of $\rm J_{21}$, it boosts the formation of $\rm H_{2}$ and cools the gas down to a few hundred K. Consequently, it elevates the critical threshold of the UV flux by a factor of 5. A similar result was found by \cite{2011MNRAS.416.2748I} but in their case the X-ray heating becomes already important for $J_{\rm X,21} = 0.01$ as they used the X-ray photo-ionization cross-section of He from \cite{Osterbrock1989}. We here use the recent cross-section by \cite{1996ApJ...465..487V} which is three orders of magnitude smaller and requires a higher value of $J_{\rm X,21}$ to elevate the value of  $J_{21}^{\rm crit}$.

\subsection{Results from 3D simulations}

In all, we have performed 25 simulations to compute the value of $J_{21}^{\rm crit}$ for the realistic Pop II spectra. We further compute the dependence of $J_{21}^{\rm crit}$ on radiation spectra, on the impact of X-ray ionization, and on variations from halo to halo. We have selected three primordial halos of a few times $\rm 10^{7}$~M$_{\odot}$ forming at $z>10 $ and varied the strength of the incident flux for a given radiation temperature.

\subsection{Impact of radiation spectra}

We assessed the dependence of $J_{21}^{\rm crit}$ on radiation spectra by selecting three representative radiation spectra i.e. $T_{\rm rad}= 2 \times 10^4, 4 \times 10^4, 8 \times 10^4 $ K for the halo A. In figure \ref{fig3}, we show the radial profiles of gas density, temperature, the $\rm H_{2}$ abundance and electron fraction. In the presence of a strong UV flux, initially the  $\rm H_{2}$ cooling remains suppressed until the halo reaches its virial temperature and a density of about $\rm 10-100~cm^{-3}$. Such trend is observed for all values of $\rm J_{21}$ irrespective of the radiation temperature.

As the halo begins to collapse the $\rm H_{2}$ abundance increases and further gets boosted in the weaker UV flux cases. For the stronger radiation fields, the formation of $\rm H_{2}$ gets delayed to higher densities until the $\rm H_{2}$ abundance reaches above $\rm 10^{-4}$ in the center of the halo and the $\rm H_{2}$ self-shielding becomes effective. By further increasing the strength of UV flux, the thermal evolution becomes isothermal and the formation of $\rm H_{2}$ remains suppressed in the presence of intense radiation fields. The $\rm H_{2}$ abundance in such cases is $\rm < 10^{-7}$ (well above the universal abundance, i.e. $\rm 10^{-3}$) and the central temperature of the halo decreases down to about 1000 K. The degree of ionization in the core of halos is higher for higher central temperatures. The density profiles for the atomic and $\rm H_{2}$ cooling  are significantly different from each other because of the different thermal evolutions. A similar trend is observed for all the radiation spectra studied here. The values of $J_{21}^{\rm crit}$ for $T_{\rm rad}= 2 \times 10^4,~4 \times 10^4 ~\& ~8 \times 10^4 $ K are $\rm 2 \times 10^4, 3 \times 10^4$ \& $\rm 3 \times 10^4$, respectively. 

For $T_{\rm rad} \geq 2 \times 10^4$ K, the variations in $J_{21}^{\rm crit}$  are not significant and the $J_{21}^{\rm crit}$ becomes constant for radiation temperatures above $\rm ~3 \times 10^4$ K. Overall, the value of $J_{21}^{\rm crit}$ is about two orders of magnitude larger than in the one-zone calculations. The differences between the one-zone and 3D simulations are also observed in \cite{Shang10} \& \cite{2014MNRAS.443.1979L}, and come from the inability of one-zone models to simulate shocks and hydrodynamical effects. These processes significantly enhance the ionization degree in realistic simulations, and stimulate the formation of $\rm H_{2}$. As a results, a stronger radiation field is required to balance this effect.

\subsection{ Variations from halo to halo}

To study the variations of $J_{21}^{\rm crit}$ from halo to halo, we selected three distinct halos and determined $J_{21}^{\rm crit}$ by varying the strength of the incident UV flux. For this purpose, we fixed the temperature of radiation spectra to $T_{\rm rad}=2 \times 10^{4}$~K. This choice is justified from the fact that the critical value only weakly depends on the radiation spectrum above this value as found in the previous section. The thermal evolution, the density profiles, the $\rm H_2$ abundances and electron fractions for these halos are shown in figure \ref{fig4}. It is noted that by increasing the intensity of the UV field, the formation of $\rm H_{2}$ remains suppressed and thermal becomes isothermal for $J_{21}^{\rm crit}$ few times $\rm 10^4$. The values of $J_{21}^{\rm crit}$ for the simulated halos are $\rm 2 \times 10^4, 4 \times 10^4$ \& $\rm 5 \times 10^4$ and also listed in table \ref{table1}. These variations from halo to halo are within a factor of 3 and are due to the differences in the density structures, spins, formation histories and the collapse redshifts. A similar dependence of $J_{21}^{\rm crit}$ on the halo properties was observed in previous studies \citep{Shang10,2014MNRAS.443.1979L}.

In figure \ref{fig41}, we show the time evolution of the density, temperature, $\rm H_{2}$ and $\rm e^{-}$ fractions for two representative cases of atomic and molecular hydrogen cooling. They are shown for the halo A and radiation spectra of $T_{\rm rad}=2 \times 10^{4}$~K. It is found that at low densities the formation of $\rm H_2$ remains suppressed and gas is heated up to $\rm \geq 10^4$ K which corresponds to the virial temperature of the halo. The halo starts to virialize at $z \sim 12$. For $J_{21}\rm = 1000$, the significant amount of $\rm H_2$ forms at densities above $\rm 10^4~cm^{-3}$ and lowers the temperature of the halo down to a few hundred K. While for $J_{21}\rm =4 \times 10^4$, the formation of $\rm H_2$ remains suppressed and leads to an isothermal collapse.

\subsection{Impact of X-ray ionization}

We investigated the impact of X-ray ionization on three different halos already exposed to a strong UV flux. To achieve this goal, we selected the UV fluxes above the critical value to see the potential influence of X-ray ionization in enhancing $J_{21}^{\rm crit}$ for a fixed radiation temperature of $T_{\rm rad}=2 \times 10^{4}$~K. An X-ray background flux of strengths $\rm  J_{X,21}=0.01~\& ~0.1$ was turned on at redshift 30. In figure \ref{fig5}, we show the profiles of density, temperature, the $\rm H_2$ and $\rm e^-$ fractions. They are very similar to the isothermal cases except for halo A. The value of $J_{21}^{\rm crit}$ is elevated by a factor of two for halo A from $\rm 2 \times 10^4$ to $\rm 4 \times 10^4$ for $J_{\rm X,21}=0.1$ while for the other two halos the impact of X-ray ionization is negligible and $J_{21}^{\rm crit}$ remains unchanged. For even higher strength of $J_{\rm X,21}=1$, the value of $J_{21}^{\rm crit}$ is enhanced by a factor of four compared to $\rm  J_{X,21}=0.01$. This comes from the fact that X-ray heating at low densities enhances the degree of ionization which further boosts the $\rm H_{2}$ fraction consequently a higher value of UV flux is required to quench the formation of molecular hydrogen.  For high values of $\rm  J_{X,21}$, our results are thus roughly consistent with the trend $J_{21}^{\rm crit} \sim J_{X,21}^{1/2}$ suggested by \cite{Inayoshi2011}, even though the influence is reduced for more moderate values.

The observations of the unresolved cosmic X-ray background at 0.2-8 keV from the Chandra Deep Fields found an intensity value of  $J_{\rm X,21}=10^{-5}$ \citep{2006ApJ...645...95H}. Here, we studied rather extreme cases a few orders of magnitude above the expected CXB \citep{2007MNRAS.374..761S,2010MNRAS.401.2635C} but the impact is still weak and is within the range of variations from halo to halo. In one-zone calculations, small variations in the degree of ionization are significant to elevate the critical value by a factor of few. On the other hand, in 3D simulations the effect of X-ray ionization is comparable to the presence of strong UV field and accretion shocks which already boost the degree of ionization. Therefore, overall the impact of X-ray ionization on $J_{21}^{\rm crit}$ is negligible for most cases studied here.

So far, we have assumed that the X-ray background is emitted from the first galaxies. In fact, the X-ray luminosity in 2-10 keV range is related to the star formation rate \citep{2003MNRAS.340..210G} as:
\begin{equation}
L_{\rm X}= 6 \times 10^{39} \left( {SFR \over 1~M_{\odot}/yr} \right)~ \mathrm{ erg~s^{-1}}
\label{eqx2}
\end{equation}
where SFR is the star formation rate. For a given SFR and distance from the source, the X-ray flux can be computed as \citep{Inayoshi2011}:
\begin{equation}
J_{\rm X,21}= 4.5 \times 10^{-3}  \left( {d \over 10~kpc} \right)^{-2} \left( {SFR \over 20~M_{\odot}yr^{-1}} \right)
\label{eqx2}
\end{equation}
This empirical relation suggests that for SFRs of 20-100 M$_{\odot}~yr^{-1}$, the X-ray flux at a distance of 10 kpc from the source varies from a few times $\rm 10^{-3}-10^{-2}$ in units of $J_{X,21}$. In the case of a nearby accreting supermassive black hole of $\rm 10^9$~M$_{\odot}$, the X-ray luminosity is about $\rm 10^{44}~erg~s^{-1}$ for a peak emission around 2 keV. This yields an X-ray flux of strength $J_{\rm X,21}=3$ at a distance of 10 kpc. These estimates suggest that the values of the X-ray background investigated here are a few orders of magnitude larger than the CXB. However, the cases investigated here may occur in the vicinity of a supermassive black hole which would therefore tend to suppress the formation of additional nearby black holes.

% Particularly, the cases studied here are expected to occur in the close vicinity of star-burst galaxies with SFR of 100-1000 M$_{\odot}/yr$. 

Recently, \cite{2014arXiv1407.1847H} have investigated the impact of only X-ray feedback (no UV flux) on a minihalo of $\rm 10^5$ M$_{\odot}$ at $z=25$ and found that X-ray feedback has a dual effect. X-ray heating is dominant at low densities, i.e. $\rm < ~1~cm^{-3}$ and the cooling ability of the gas gets enhanced above $\rm 10^2~cm^{-3}$ due to the higher degree of ionization. They found that X-ray feedback has a so-called Goldilocks range of the X-ray background for which it promotes star formation while for the higher strengths it suppresses the collapse. This Goldilocks range of CXB starts around $J_{\rm X,21}=0.1$ and complete blowout occurs for $J_{\rm X,21}=10$ where it completely suppresses the collapse of a minihalo. 

The halos studied here are about an order of magnitude more massive than the minihalos therefore the impact of X-rays is expected to be less severe. In figure \ref{fig51}, we show the impact of X-ray heating and ionization for various strengths of $J_{\rm X,21}$ from 0.01-10. The temperature of the cold component of the gas scales with the strength of X-ray feedback. It is noted that for the highest strength of $J_{\rm X,21}$ the gas is heated above $10^4$ K and cools by Lyman alpha radiation. X-rays increase the degree of ionization, which consequently promotes the cold component of the gas at higher densities. This is particularly notable for $J_{\rm X,21}=10$ for which the central temperature of the halo reaches about 300 K due to the higher $\rm H_{2}$ abundance. In a nutshell, for the X-ray background to be effective to raise the critical threshold for UV flux it should have at least a strength of $J_{\rm X,21} \geq 0.1$.  

We further explored the influence of X-ray feedback of strength $J_{\rm X,21}=100$ for the halo A. We found that the collapse of the halo is delayed by $\Delta~z=1$, i.e., 50 million years, but it is not disrupted. Therefore, blowout of massive primordial halos is very unlikely and may occur only for very extreme X-ray feedback of strength higher than investigated here. It is noted that such an extreme strength of the X-ray background can only be achieved at a distance of 1 kpc from an accreting supermassive black holes of a billion solar masses. The impact of X-ray feedback on the collapse dynamics and subsequent star formation in massive hales should be investigated in future studies via 3D simulations.

\subsection{Implications for the formation of DCBHs}

It is a prerequisite for DCBHs that the formation of $\rm H_2$ remains suppressed in their ``potential embryos'' atomic cooling halos. Quenching the $\rm H_2$ formation mandates the presence of an intense UV flux which can be achieved in the close proximity (about 10 kpc) of a star forming galaxy \citep{Dijkstra08,Agarwal12,Dijkstra14}. It is likely that such a flux is provided by Pop II stars which are abundant enough to illuminate the number of DCBHs sites comparable with the observed number density of high redshift quasars at $z>6$ \citep{Dijkstra14,2014MNRAS.443.1979L}.

The accurate determination of $J_{21}^{\rm crit}$ for realistic Pop II spectra is highly relevant to predict the sites of DCBHs and to test the feasibility of the direct collapse scenario.   \cite{Agarwal12}, and \cite{Dijkstra14} have estimated the abundance of DCBHs from semi-analytical calculations along with N-body and Monte-Carlo simulations. They assumed that DCBHs are formed in the atomic cooling halos irradiated by $J_{21}> J_{21}^{\rm crit}$ by taking $J_{21}^{\rm crit}=30$ and $J_{21}^{\rm crit}=300$ respectively. Their predictions lead to the DCBHs abundances of $\rm 3 \times 10^{-3}/cMpc^{3}$ and $\rm 10^{-7}/cMpc^{3}$ (respectively) higher than the observed SMBHs number density of $\rm 10^{-9}/cMpc^{3}$ \citep{2003AAS...20312901F,2013ApJ...779...24V}. Recently, \cite{Omukai2014} extrapolated the estimates of \cite{Dijkstra14} by using their one-zone estimates of $J_{21}^{\rm crit}=1400$ and found that number density of DCBHs drops to $\rm 10^{-10}/cMpc^{3}$ at $z=10$.

\begin{figure}
%  \vspace{-4.0cm}
\hspace{-6.0cm}
\centering
\begin{tabular}{c}
\begin{minipage}{4cm}
\includegraphics[scale=0.5]{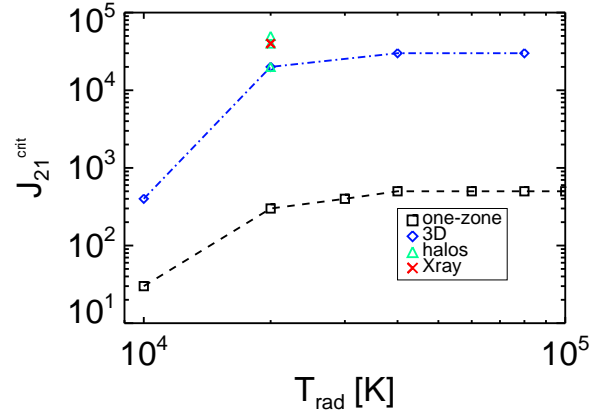}
\end{minipage}
\end{tabular}
\caption{Summary plot. In this figure, we show the results from one-zone model and 3D simulations performed to compute the variations from halo to halo, dependence on the radiation spectra \& the impact of X-ray ionization.}
\label{fig6}
\end{figure}

In figure \ref{fig6}, we summarize our estimates of $J_{21}^{\rm crit}$ from one-zone, 3D simulations, variations from halo to halo, dependence on the radiation spectra and from the impact of X-ray ionization. Our calculation of $J_{21}^{\rm crit}$ for the realistic Pop II spectra suggest that $J_{21}^{\rm crit}$ varies from 20,000-50,000. Extrapolating the results of \cite{Dijkstra14} with $J_{21}^{\rm crit}=40,000$ the number density of DCBH sites drops about 5 orders of magnitude below the observed quasars density. Our results suggest that direct collapse black holes forming through an isothermal may be even more rare than previously perceived. However, as recently shown by \cite{Latif2014ApJ}, the formation of massive objects is still conceivable for moderate amounts of $\rm H_2$ cooling, implying that this formation channel deserves further exploration in the future. 

\section{Discussion}

The prime requisitions for the formation of isothermal direct collapse black holes are that the hosting halos must be of primordial composition and remain $\rm H_{2}$ free. The first condition can be accomplished in the massive halos forming from primordial gas composition at $z=15$. The suppression of $\rm H_{2}$ demands the presence of a strong UV flux during their formation. Such flux is conceived to be produced by the second generation of stars at earlier cosmic epochs. In this study, we compute the critical value of UV flux necessary for the formation of direct collapse black holes for the realistic Pop II spectra by preforming high resolution cosmological simulations. We further compute the dependence of $J_{21}^{\rm crit}$ on the radiation spectra, study the impact of X-ray ionization and variations for halo to halo. 

In these simulations, we followed the initial collapse of DM halos to high densities from which stars and black holes may subsequently form. The masses of these stars depend on the thermal properties of the host halos and massive to supermassive stars are expected to form. Particularly, for the isothermal collapse, a supermassive star of $\rm 10^5~M_{\odot}$ is expected to form which will later collapse into a BH. The X-ray feedback from such a direct collapse BH may trigger in-situ star formation in about 0.5 Myrs after its formation \citep{2014arXiv1409.0543A}. The local star formation may induce metal enrichment and consequently change the thermal structure of the gas, and also the gas dynamics due to the injection of turbulence energy. Overall, we do not expect star formation to inhibit black hole growth, and in fact the injected turbulent energy helps to transfer the angular momentum and maintain the accretion rates \citep{2013A&A...560A..34W}.

Our findings suggest that the value of $J_{21}^{\rm crit}$ for realistic Pop II spectra is a few times $\rm 10^4$. For the simulated halos, it varies from $\rm 2 \times 10^{4}-5 \times 10^{4}$ and also weakly depends on the radiation spectra in the range $T_{\rm rad}=2 \times 10^4-10^{5}$ K. We further found that the impact of X-ray ionization for the halos irradiated by strong UV flux is negligible for $J_{\rm X,21} \leq 0.1$ and does not elevate the $J_{21}^{\rm crit}$ for UV flux in contrary to the previous studies. For $J_{\rm X,21} > 0.1$ it does have a factor of a few effect on $J_{21}^{\rm crit}$. However, we noted that in one of the simulated halos it changed the $J_{21}^{\rm crit}$ from $\rm 2 \times 10^{4}- 4 \times 10^{4}$. This increase is within the expected scatter in the values of $J_{21}^{\rm crit}$. The results from 3D simulations differ by about two orders of magnitude from one-zone calculations due to the inability of one-zone models to simulate shocks, collapse dynamics and hydrodynamical effects as found in the previous studies \citep{Shang10,2014MNRAS.443.1979L}. In comparison with earlier works, our one-zone results differ by a factor of 3 from \cite{Omukai2014}. This difference arises by considering the dissociative tunneling effect in the collisional dissociation of $\rm H_{2}$. For very strong X-ray fluxes, we can roughly confirm their trend of $J_{21}^{\rm crit} \sim  J_{X,21}^{1/2}$, even though such cases may occur in the presence of a supermassive black hole.

For this study, we have presumed that the halo is illuminated by an isotropic background UV flux of constant strength. It is likely that a direct collapse halo is irradiated by multiple sources, in such case the expected background is approximately isotropic as seen in the simulations of \cite{2014MNRAS.443..648A}. However, the recent work by \cite{2014arXiv1407.4472R} shows that in the case of an anisotropic single source the value of $J_{21}^{\rm crit}$ is few times $\rm 10^3$ and almost consistent with our results. It may also be noted that \cite{2014arXiv1407.4472R} consider a monochromatic source of Lyman-Werner radiation and ignore other dissociation channels particularly $\rm H^{-}$. We further suggest that our estimates of $J_{21}^{\rm crit}$ are also valid both for pop III stars with radiation temperature of $\rm 10^5$ K. We have not considered here the impact of ionizing UV radiation which further enhances the degree of ionization and elevates the value of $J_{21}^{\rm crit}$ \citep{2014arXiv1405.2081J}. It should also be noted that strong dissipation of magnetic energy into thermal energy may reduce the value of $J_{21}^{\rm crit}$ \citep{2010ApJ...721..615S,2013A&A...553L...9V}. Once the gas is dense enough about $\rm 10^{4}~cm^{-3}$ and above temperature of few thousand K then collisional dissociation becomes effective leading to the destruction of $\rm H_{2}$ \citep{2012MNRAS.422.2539I}.

The higher value of $J_{21}^{\rm crit}$ (i.e. $\rm 10^{4}$) found from this work suggests that complete photo-dissociation sites are very rare and less abundant than the number of quasars observed at $z=10$. However, as found in \cite{Latif2014ApJ} even in the cases where $\rm H_2$ is not completely dissociated still massive objects of $\rm 10^4$~M$_{\odot}$ may form provided that accretion rates remain higher than $\rm 0.1$ M$_{\odot}~yr^{-1}$. For such accretion rates, the feedback from the protostar remains suppressed and accretion can proceed for longer time scales \citep{Hosokawa13,Schleicher13}. Even if fragmentation occurs at high densities inside the disk due to the $\rm H^-$ or dust cooling, clumps may merge in the center on a shorter time scale than the Kelvin-Helmholtz time scale \citep{2014arXiv1406.5058I}. Such seeds are still an order of magnitude more massive than remnants of Pop III stars or stellar cluster. We argue here that such sites are strong contenders to the scarce isothermal collapse.

Our study does not capture the subtle time dependent effects of Pop II spectra such as spectral dependence on the age, metallicity and stellar mass function. Recent attempts to model these effects are in disagreement \citep{2014arXiv1407.4115A,Omukai2014}, therefore further investigations in this direction are required. In the future, alternative channels for the formation of seed black holes should be explored to test their feasibility.

\section*{Acknowledgments}
The simulations described in this work were performed using the Enzo code, developed by the Laboratory for Computational Astrophysics at the University of California in San Diego (http://lca.ucsd.edu). We are grateful to Philipp Stancil and Bob Forrey for sharing the H$^-$ cross-sections. We thank Kohei Inayoshi for his valuable feedback on X-ray physics. We also thank Daniele Galli for fruitful discussions on H$_2$ photodissociation. We acknowledge research funding by Deutsche Forschungsgemeinschaft (DFG) under grant SFB $\rm 963/1$ (project A12) and computing time from HLRN under project nip00029. DRGS and SB thank the DFG for funding via the Schwerpunktprogram SPP 1573 ``Physics of the Interstellar Medium'' (grant SCHL $\rm 1964/1-1$). TG acknowledges the Centre for Star and Planet Formation funded by the Danish National Research Foundation. The simulation results are analyzed using the visualization toolkit for astrophysical data YT \citep{2011ApJS..192....9T}.

\bibliography{blackholes.bib}
 
\newpage

\appendix

\section{}

\subsection{Chemical Model}
The details of our chemical model are described in the following subsections. The chemical reactions and their rate coefficients are listed in table 1.
\subsection{Photochemistry}
For the photo-dissociation of $\rm H_{2}$, we use the rate provided by \citet{GloverJappsen2007} and multiply it with the Lyman Werner flux centered at $h\overline{\nu}$=12.87 eV normalized by the spectrum at Lyman limit as the absorption of Lyman Werner photons occurs in the narrow frequency range $12.24 {\rm eV}< h$ $\nu$ $ < 13.51$ eV \citep{Abel97,GloverJappsen2007}. It is further re-scaled by $J_{21}$
\begin{equation}\label{eq:integration}
	k_{\rm H_{2}}[T_{\rm rad}] = 1.38\times10^{-12} J_{21} \frac{B(12.87~\mathrm{eV}, T_{\rm rad})}{B(13.6 ~\mathrm{eV}, T_{\rm rad})} ~~~~~~~~~~[s^{-1}]
\end{equation}
here $B(E, T_{\rm rad})$ is the black-body spectrum and is defined as:
\begin{equation}
		B(E, T_{\rm rad}) = \frac{2 E^3}{h^2 c^2}\frac{1}{(e^{E/k_bT_{\rm rad}}-1)}
\end{equation}
where $h$ is the Planck constant, $c$ the speed of light, and $k_b$ the Boltzmann constant. The rate is computed by assuming that H$_2$ molecule is in the roto-vibrational ground state ($\nu = 0, J = 0$). However, relaxing this assumption does not strongly affect the final rate \citep{GloverJappsen2007}.

For the H$^-$ photo-detachment cross-section, we use recent data which includes the absorption of the H$^-$ resonance states near 11 eV (Stancil P. private communication, similar to \citealt{Miyake2010}). These contributions were neglected in previous calculation of the rate by \citealt{Wishart1979}. We note that by taking into account this contribution does not change the final rate significantly. The H$^-$ photo-detachment rate is evaluated by integrating the cross-section over different radiation black-body spectra normalized at 13.6 eV in the following way:
\begin{equation}
		k_{\rm H^{-}}[T_{\rm rad}] = 6.2415 \times 10^{-10} J_{21} \frac{\frac {4 \pi} {h} \int_{0.76}^{13.6} \frac{ \sigma(E)}{E} B(E, T_{\rm rad})dE}{B(13.6 ~\mathrm{eV}, T_{\rm rad})}
\end{equation}
The photo-dissociation of $\rm H_{2}^{+}$ is computed in a similar way and the cross-section is reported in table \ref{tab:rates}.

\subsection{X-ray physics}
The impact of X-ray heating and ionization on hydrogen and helium atoms has been investigated in many studies \citep{Shull1985,Wolfire1995,Ricotti2002,Ricotti2004,2005A&A...436..397M, Ferrara2008,Furlanetto2010}. The primary photo-ionization produces energetic electrons which can subsequently cause a secondary ionization of atoms. The total (primary + secondary) photo ionization rate is
\begin{equation}\label{equation1}
        \zeta^{i}_{tot}  = \zeta^{i}_p + \sum_{j=\rm{H, He}} \frac{n_{j}}{n_{i}}\zeta^j_p
       \langle\phi^i\rangle\\
%        \zeta^{He}_{tot} = \zeta^{He}_p + \sum_{i=\rm{H, He}}\zeta^i_p\phi^{He}\\
\end{equation}
\noindent where the index $p$ stands for the primary, $n$ is the number density of the given specie, $i$ represents H and He. The second term on the right hand side shows the contribution of secondary ionization. The primary photo-ionization rate $\zeta^i_p$ is
\begin{equation}
        \zeta^i_p = \frac {4 \pi} {h} \int{\frac{J_X(E)}{E} e^{-\tau(E)}\sigma^i(E)dE}.\\
\end{equation}
The flux $J_X(E)$ is given in equation \ref{eqx}, $\sigma^i$ are the  photo-ionization cross sections of H and He and are taken from \citet{Verner1996}. It is worth noting that the cross section for He given by \cite{Osterbrock1989} and employed in \citet{Inayoshi2011} is few orders of magnitude higher. This results in higher photo-heating/photo-ionization and reduces the value of $J_{\rm X,21}$ to become effective for $J_{21}^{\rm crit}$ as already discussed in the results section.

%  and expressed in terms of $J_{21,X}$ erg cm$^{-2}$ s$^{-1}$ sr$^{-1}$ Hz$^{-1}$
% \begin{equation}
% 	J_X(E) = J_{21,X} \times 10^{-21} (E/E_0)^{-1.5}
% \end{equation}
% where we have introduced a subscript "$X$" to differentiate from the standard  UV $J_{21}$.
Here $\tau$ is the opacity expressed as \citep{Inayoshi2011}
\begin{equation}
        \tau(E) = \sum_{i=\rm{H,He}}\sigma^i(E) N_i
\end{equation}
$ N_i$ is the column density defined as
\begin{equation}
        N_{i} = n_{i} \lambda_J
\end{equation}
where $\lambda_j$ is the Jeans length defined as
\begin{equation}
	\lambda_J = \sqrt{\frac{\pi k_b T}{G\rho\mu m_H}}
\end{equation}
$G$ the gravitational constant, $m_H$ the proton mass, $\rho$ the total mass density, and $\mu$ the mean molecular weight evaluated as 
\begin{equation}
	\mu = \frac{\sum_k\rho_k}{\rho}\\
\end{equation}
here $k$ represents all the species. 

The number of secondary ionization of H and He per primary ionization, $\phi^H$ and $\phi^{He}$ are taken from \citet{Shull1985} and are defined as:
\begin{equation}\label{secq1}
        \phi^H (E,x_e) = \left(\frac{E}{13.6 ~\mathrm{eV}} - 1\right) 0.3908(1-x_e^{0.4092})^{1.7592}
\end{equation} 
\begin{equation}\label{secq2}
        \phi^{He} (E,x_e) = \left(\frac{E}{24.6 ~\mathrm{eV}} - 1\right) 0.0554(1-x_e^{0.4614})^{1.666}
\end{equation}
where $x_e$ is the electron fraction.

The above quantities are then averaged over the X-rays spectrum as
\begin{equation}
	\langle\phi^i\rangle = \frac{\int{J_X(E) \phi^i(E,x_e)dE}}{\int{J_X(E)}dE}
\end{equation}
and are used in eq. \ref{equation1}.
The fitting formulas for the secondary ionization given in equations \ref{secq1} \& \ref{secq2} are valid for energy $ E > 100$ eV and for a gas mixture of H and He.
% For the applications considering lower energies different fitting formulae other than \citet{Shull1985} should be used, as for instance the one discussed in the appendix of \citet{Wolfire1995} and employed by \citet{Inayoshi2011}.

The photo-ionization heating is  given by
\begin{equation}
        \Gamma = \Gamma^{\rm H} + \Gamma^{\rm He}
\end{equation}
where $\Gamma^i$ is defined as following
\begin{equation}
        \Gamma^i = \frac {4 \pi} {h}\int{\frac{J_X(E)}{E} e^{-\tau(E)}\sigma^i(E)E^i_h(E,x_e)dE}
\end{equation}
with $ E_h^i(E,x_e)$ is the fraction of primary electron energy which goes into heating \citep{Shull1985}. The heating is given in units of eV~s$^{-1}$.
\begin{equation}
       E^i_h(E,x_e) = (E-E^i_0)0.9971[1 - (1-x_e^{0.2663})^{1.3163}]
\end{equation}
with $ i =$ H, He, $E_0^{\rm H} = 13.6$ eV, and $E_0^{\rm He} = 24.6$ eV. 

Rates \& reactions used in this study are listed in table below.

\appendix
\begin{table*}
        \caption{The list of reactions and rates included in our chemical network. Here $T$ is the gas temperature in K while $T_e$ is the gas temperature in eV.}\label{tab:rates}
        \begin{tabular}{@{}lllc}
                \hline\hline
                Reaction & Rate coefficient (cm$^3$ s$^{-1}$) & & Ref.\\
                \hline
 (1)       H + e$^-$ $\rightarrow$ H$^+$ + 2e$^-$  & $k_1$ = exp[-32.71396786+13.5365560 ln $T_e$ & & 1 \\
        & - 5.73932875 (ln $T_e$)$^2$+1.56315498 (ln $T_e$)$^3$ & &  \\
        & - 0.28770560 (ln $T_e$)$^4$+3.48255977 $\times$ 10$^{-2}$(ln $T_e$)$^5$ & & \\
        & - 2.63197617 $\times$ 10$^{-3}$(ln $T_e$)$^6$+1.11954395 $\times$ 10$^{-4}$(ln $T_e$)$^7$ & \\
        & - 2.03914985 $\times$ 10$^{-6}$(ln $T_e$)$^8$] & \\
(2)        H$^+$ + e$^-$ $\rightarrow$ H  + $\gamma$ & $k_2$ = 3.92 $\times$ 10$^{-13}$ $T_e$ $^{-0.6353}$ & $T \le 5500$ K & 2 \\
        & $k_2$ = $\exp$[-28.61303380689232 & $T > 5500$ K & \\
& - 7.241 125 657 826 851 $\times$ 10$^{-1}$ ln $T_e$\\
& - 2.026 044 731 984 691 $\times$ 10$^{-2}$ (ln $T_e$)$^2$\\
& - 2.380 861 877 349 834 $\times$ 10$^{-3}$ (ln $T_e$)$^3$\\
& - 3.212 605 213 188 796 $\times$ 10$^{-4}$ (ln $T_e$)$^4$\\
& - 1.421 502 914 054 107 $\times$ 10$^{-5}$ (ln $T_e$)$^5$\\
& + 4.989 108 920 299 510  $\times$ 10$^{-6}$ (ln $T_e$)$^6$\\
& + 5.755 614 137 575 750  $\times$ 10$^{-7}$ (ln $T_e$)$^7$\\
& - 1.856 767 039 775 260  $\times$ 10$^{-8}$  (ln $T_e$)$^8$\\
& - 3.071 135 243 196 590  $\times$ 10$^{-9}$  (ln $T_e$)$^9$]  \\
(3)       He + e$^-$ $\rightarrow$ He$^+$ + 2e$^-$  & $k_3$ = $\exp$[-44.09864886 & $T_e > 0.8$ eV & 1 \\
        & + 23.915 965 63 ln$T_e$ \\
        & - 10.753 230 2 (ln $T_e$)$^2$\\
        & + 3.058 038 75 (ln $T_e$)$^3$\\
        & - 5.685 118 9 $\times$ 10$^{-1}$ (ln $T_e$)$^4$\\
        & + 6.795 391 23 $\times$ 10$^{-2}$ (ln $T_e$)$^5$\\
        & - 5.009 056 10 $\times$ 10$^{-3}$ (ln $T_e$)$^6$\\
        & + 2.067 236 16 $\times$ 10$^{-4}$ (ln$T_e$)$^7$\\
        & - 3.649 161 41 $\times$ 10$^{-6}$ (ln $T_e$)$^8$]  & \\
(4) He$^+$ + e$^-$ $\rightarrow$ He + $\gamma$ & $k_4$ =  3.92 $\times$ 10$^{-13}$ $T_e$ $^{-0.6353}$ &  $T_e \le 0.8$ eV & 3\\
        & $k_4 = $ + 3.92 $\times$ 10$^{-13}$ $T_e^{-0.6353}$ & $T > 0.8$ eV \\
        & + 1.54 $\times$ 10$^{-9}$ $T_e^{-1.5}$ [1.0 + 0.3 / $\exp$(8.099 328 789 667/$T_e$)]  \\
        & /[$\exp$(40.496 643 948 336 62/$T_e$)]\\
        &  & \\
 (5)       He$^+$ + e$^-$ $\rightarrow$ He$^{++}$ + 2e$^-$ & $k_5$ = $\exp$[-68.710 409 902 120 01 & $T_e > 0.8 $ eV & 4\\
        & + 43.933 476 326 35 ln$T_e$ \\
        & - 18.480 669 935 68 (ln $T_e$)$^2$ \\
        & + 4.701 626 486 759 002 (ln $T_e$)$^3$ \\
        & - 7.692 466 334 492 $\times$ 10$^{-1}$ (ln $T_e$)$^4$\\
        & + 8.113 042 097 303 $\times$ 10$^{-2}$ (ln $T_e$)$^5$\\
                & - 5.324 020 628 287 001 $\times$ 10$^{-3}$ (ln $T_e$)$^6$\\
        & + 1.975 705 312 221 $\times$ 10$^{-4}$ (ln $T_e$)$^7$ \\
        & - 3.165581065665 $\times$ 10$^{-6}$ (ln $T_e$)$^8$] & \\
(6)        He$^{++}$ + e$^-$ $\rightarrow$ He$^+$ + $\gamma$ & $k_6$ = 
        1.891 $\times$ 10$^{-10}$/$[1+
       \sqrt{T/9.37}]^{0.2476}/[ 1+ \sqrt{T/(2.774\times 10^6)}]^{1.7524}$/$\sqrt{T/9.37}$ & & 5\\
(7)        H + e $\rightarrow$ H$^-$ + $\gamma$ & $k_7$ = $1.4 \times 10^{-18}T^{0.928}\exp(-T/16200)$ & & 6\\
(8)        H$^-$ + H $\rightarrow$ H$_2$ + e$^-$ & $k_8$ = $a_1(T^{a_2}+a_3 T^{a_4}+a_5T^{a_6})/(1+a_7T^{a_8}+a_9T^{a_{10}}+a_{11}T^{a_{12}}$) & & 7\\
        & $a_1 = 10^{-10}$\\
        & $a_2 = 9.8493 \times 10^{-2}$\\                                                                       
		& 	$a_3 = 3.2852 \times 10^{-1}$\\                                                                      
		& $a_4 = 5.5610 \times 10^{-1}$\\                                                                       
		& $a_5 = 2.7710 \times 10^{-7}$\\                                                                       
		& $a_6 = 2.1826$                           \\                                            
	 	& $a_7 = 6.1910 \times 10^{-3}$\\                                                                       
		& $a_8 = 1.0461$\\                                                                       
		& $a_9 = 8.9712 \times 10^{-11}$\\                                                                       
		& $a_{10} = 3.0424$\\                                                                      
		& $a_{11} = 3.2576 \times 10^{-14}$\\                                                                      
		& $a_{12} = 3.7741$\\
(9)        H + H$^+$ $\rightarrow$ H$_2^+$ + $\gamma$ & $k_9$ =    		
        $2.10\times 10^{-20}(T/30)^{-0.15}$ & $T < 30 $ K &  8\\
        & dex$[-18.20-3.194\log_{10}T+1.786(\log_{10} T)^{2}-0.2072(
        \log_{10}T)^3]$ & $T \geq 30$ K  \\ 
        \hline
\end{tabular}
\\
1: \cite{Janev1987}, 2:  \citet{Abel97} fit by data from \citet{Ferland1992}, 3: \citet{Cen92,Aldrovandi1973} \\
4: Aladdin database, see \citet{Abel97}, 5: \citet{Verner1996}, 6: \citet{DeJong1972}, 7: \citet{Kreckel2010} \\
8: \citet{Coppola2011}
\end{table*}

\begin{table*}
        \contcaption{The list of reactions and rates included in our chemical network. Here $T$ is the gas temperature in K while $T_e$ is the gas temperature in eV.}
        \begin{tabular}{@{}lllc}
                \hline\hline
                Reaction & Rate coefficient (cm$^3$ s$^{-1}$) & & Ref.\\
                \hline
(10)       H$_2^+$ + H $\rightarrow$ H$_2$ + H$^+$ & $k_{10}$ = 6.0 $\times$ 10$^{-10}$ & & 9\\
(11)       H$_2$ + H$^+$ $\rightarrow$ H$_2^+$ + H & $k_{11}$ = 	   		
        [$b_1+b_2\ln(T)+b_3(\ln(T))^2+b_4(\ln(T))^3+b_5(\ln(T))^4+b_6(\ln(T))^5$ & & 10\\
        & $+b_7(\ln(T))^6+b_8(\ln(T))^7]\exp(-\mathrm{a}/T)$\\
        & $a = 2.1237150 \times 10^4$\\
		& $b_1=-3.3232183 \times 10^{-7}$\\
		& $b_2= 3.3735382 \times 10^{-7}$\\
		& $b_3=-1.4491368 \times 10^{-7}$\\
		& $b_4= 3.4172805 \times 10^{-8}$\\
		& $b_5=-4.7813728 \times 10^{-9}$\\
		& $b_6= 3.9731542 \times 10^{-10}$\\
		& $b_7=-1.8171411 \times 10^{-11}$\\
		& $b_8= 3.5311932 \times 10^{-13}$\\
(12)        H$^-$ + e$^-$ $\rightarrow$ H + 2e$^-$  & $k_{12}$ = $\exp$[-18.018 493 342 73 & & 1 \\
        & + 2.360 852 208 681 ln$T_e$\\
        & - 2.827 443 061 704 $\times$ 10$^{-1}$ (ln $T_e$)$^2$\\
        & + 1.623 316 639 567 $\times$ 10$^{-2}$ (ln $T_e$)$^3$\\
        & - 3.365 012 031 362 999 $\times$ 10$^{-2}$ (ln $T_e$)$^4$\\
        & + 1.178 329 782 711  $\times$ 10$^{-2}$ (ln $T_e$)$^5$\\
        & - 1.656 194 699 504  $\times$ 10$^{-3}$ (ln $T_e$)$^6$\\
        & + 1.068 275 202 678  $\times$ 10$^{-4}$ (ln $T_e$)$^7$\\
        & - 2.631 285 809 207  $\times$ 10$^{-6}$ (ln $T_e$)$^8$]& \\
(13)        H$^-$ + H $\rightarrow$ 2H +  e$^-$ & $k_{13}$ = 2.56 $\times$ 10$^{-9}$ $T_e^{1.78186}$ & $T_e \le  0.04$ eV  & 11\\
        & $k_{13}$ =  $\exp$[-20.372 608 965 333 24 & $T_e >  0.04$ eV \\
        & + 1.139 449 335 841 631 ln $T_e$ \\
        & - 1.421 013 521 554 148 $\times$ 10$^{-1}$ (ln $T_e$)$^2$\\
        & + 8.464 455 386 63 $\times$ 10$^{-3}$ (ln $T_e$)$^3$\\
        & - 1.432 764 121 299 2 $\times$ 10$^{-3}$ (ln $T_e$)$^4$\\
        & +2.012 250 284 791 $\times$ 10$^{-4}$ (ln $T_e$)$^5$\\
        & + 8.663 963 243 09 $\times$ 10$^{-5}$ (ln $T_e$)$^6$\\
        & - 2.585 009 680 264 $\times$ 10$^{-5}$ (ln $T_e$)$^7$\\
        & + 2.455 501 197 039 2 $\times$ 10$^{-6}$ (ln $T_e$)$^8$\\
        & - 8.068 382 461 18 $\times$ 10$^{-8}$ (ln $T_e$)$^9$]\\
 (14)       H$^-$ + H$^+$ $\rightarrow$ 2H + $\gamma$ & $k_{14}$ = $2.96 \times 10^{-6}/\sqrt{T}-1.73\times10^{-9}+2.50\times 10^{-10}\sqrt{T}-7.77\times10^{-13}T$ & & 12\\
(15)        H$^-$ + H$^+$ $\rightarrow$ H$_2^+$ + e$^-$ & $k_{15}$ = 10$^{-8} \times T^{-0.4}$& & 13\\
(16)        H$_2^+$ + e $\rightarrow$ 2H + $\gamma$ & $ k_{16}$ = $10^6(4.2278\times10^{-14}-2.3088\times10^{-17}T+7.3428\times10^{-21}T^2-7.5474\times10^{-25}T^3$ & & 8\\
        &$+3.3468\times10^{-29}T^4-5.528\times10^{-34}T^5)$  \\
(17)       H$_2^+$ + H$^-$ $\rightarrow$ H + H$_2$ & $k_{17}$ = 5.0 $\times$ 10$^{-7}$ $\sqrt{10^2/T}$ & &14 \\
(18)        3H $\rightarrow$ H$_2$ + H & $k_{18}$ = $6\times10^{-32}T^{-0.25}+2\times10^{-31}T^{-0.5}$ & & 15\\
(19)        H$_2$ + H $\rightarrow$ 3H & $k_{19}$, see ref. & & 16\\
(20) H$_2$ + H + H $\rightarrow$ H$_2$ + H$_2$ & $k_{20} = k_{18}/8$ & & 17\\
(21) H$_2$ + H$_2$ $\rightarrow$ H$_2$ + H + H & $k_{21} = kh_{21}^{(1 - 
  a_{21})}kl_{21}^{a_{21}}$ & & 18\\
  &$kl_{21} =1.18\times10^{-10}\exp(-6.95\times10^4/T)$\\
  &$kh_{21} = 8.125\times10^{-8}T^{-0.5}\exp(-5.2\times10^4/T)(1-\exp(-6\times10^3/T))$\\
&$ncr_{21} = \mathrm{dex}[4.845-1.3\log_{10}(T/10^4)+1.62[\log_{10}(T/10^4)]^2]$\\
&$a_{21}=1/(1+(\mathrm{nH}/ncr_{21}))$\\
(22) He$^+$ + H $\rightarrow$ He + H$^+$ & $k_{22} = 1.20\times10^{-15}(T/300)^{0.25}$ & & 19\\
(23) He + H$^+$ $\rightarrow$ He$^+$ + H & $k_{23} = 1.26\times10^{-9}T^{-0.75}\exp(-1.275\times10^5/T)$ & $T \leq 10^4$ K & 19\\
& $k_{24} = 4\times10^{-37}T^{4.74}$ & $T > 10^4$ K \\
(24) H$_2$ + e$^-$ $\rightarrow$ H + H + e$^-$ & $k_{24} = 4.38
\times10^{-10}T^{0.35}\exp(-102000/T)$ & & 20\\
(25) H$_2$ + e$^-$ $\rightarrow$ H + H$^-$ & $k_{25} = 35.5\times T^{-2.28}
\exp(-46707/T)$ & & 21\\
(26) H$_2$ + $\gamma$ $\rightarrow$ H + H & $k_{26} = 1.38\times10^9 $ & in units of s$^{-1}$ & 22 \\
(27) H$_2^+$ + $\gamma$ $\rightarrow$ H + H$^+$ & $\sigma$ = dex$[-40.97+6.03 E-0.504 E^2+1.387\times10^{-2}E^3]$ & 2.65 eV $< E <$ 11.27 eV & 23\\
& $\sigma$ = dex$[-30.26+2.79 E-0.184 E^2+3.535\times10^{-3}E^3]$ & 11.27 eV $< E < 21.0$ eV\\
(28) H$^-$ + $\gamma$ $\rightarrow$ H + e$^-$ & private communication P. Stancil, 2014 & see text & 24\\
(29)  H + $\gamma$ $\rightarrow$ H$^+$ + e$^-$ &   fit parameters from 5 & &5\\
(30)  He + $\gamma$ $\rightarrow$ He$^+$ + e$^-$ &   fit parameters from 5 & & 5\\

  \hline       
\end{tabular}
\\
9 \citet{Karpas1979}, 10: \citet{Savin2004}, 11: \citet{Abel97} based on \citet{Janev1987}, 12: \citet{Stenrup2009}, 13: \citet{Poulaert1978},
14: \citet{Dalgarno1987}, 15: \citet{Forrey2013}, 16: \citet{Martin1996}, 17: rescaled following \citet{GloverAbel08}\\ 18: \citet{Omukai2000}, 19: \citet{Yoshida2006}, 20: \citet{Mitchell1983} fit of data by \citet{Corrigan1965}, 21: \citet{Capitelli2007}, 22: \citet{GloverJappsen2007}, 23: \citet{Shapiro1987}, originally from \citet{Dunn1968}, 24: see also \citet{Miyake2010}
\end{table*}

\end{document}